\documentclass{pasa}%

\title[Limits of NIR surveys]{Detection, size, measurement and structural analysis limits for the 2MASS, UKIDSS-LAS \& VISTA VIKING surveys}
\author[Andrews et al.]{Stephen K. Andrews$^1$, Lee S. Kelvin$^{1,2,3}$, Simon P. Driver$^{1,2}$ \and Aaron S.~G. Robotham$^{1,2}$ \\
\affil{$^1$ International Center for Radio Astronomy Research, 7 Fairway, The University of Western Australia, Crawley, Perth, Western Australia 6009, Australia}%
\affil{$^2$ School of Physics and Astronomy, University of St. Andrews, North Haugh, St. Andrews, Fife, KY16 9SS, UK}
\affil{$^3$ Institut f\"{u}r Astro- und Teilchenphysik, Universit\"{a}t Innsbruck, Technikerstra{\ss}e 25, 6020 Innsbruck, Austria}
\affil{$^*$ Email: sandrews.astro@gmail.com}}%
\jid{PASA}
\doi{10.1017/pas.\the\year.xxx}
\jyear{\the\year}

\usepackage[authoryear]{natbib}
\usepackage{comment}
\usepackage{multirow}
\usepackage{array}
\PassOptionsToPackage{hyphens}{url}\usepackage{hyperref}
\usepackage{longtable}

\newcommand{\degree}{\ensuremath{^\circ}}

\begin{document}%
\begin{abstract}
The 2MASS, UKIDSS-LAS and VISTA VIKING surveys have all now observed the GAMA 9hr region in the $K_s$ band. Here we compare the detection rates, photometry, basic size measurements, and single-component \textsc{GALFIT} structural measurements for a sample of \textbf{37,591} galaxies. We explore the sensitivity limits where the data agree for a variety of issues including: detection, star-galaxy separation, photometric measurements, size and ellipticity measurements, and S\'ersic measurements. We find that 2MASS fails to detect at least 20\% of the galaxy population within all magnitude bins, however for those that are detected we find photometry is robust ($\pm$ 0.2~mag) to 14.7~AB mag and star-galaxy separation to 14.8~AB mag. For UKIDSS-LAS we find incompleteness starts to enter at a flux limit of 18.9~AB mag, star-galaxy separation is robust to 16.3~AB mag and structural measurements are robust to 17.7~AB mag. VISTA VIKING data is complete to approximately 20.0~AB mag and structural measurements appear robust to 18.8~AB mag.
\end{abstract}
\begin{keywords}
galaxies: fundamental parameters  --- surveys --- infrared: galaxies
\end{keywords}
\maketitle%

\section{INTRODUCTION}

Many important extra-galactic measurements rely on robust measurement of the fluxes and structural properties for large samples of galaxies (e.g. \citealt{allen06}, \citealt{simard11}, \citealt{lackner12}, \citealt{kelvin12}). Examples include the bivariate brightness distribution (e.g. \citealt{smith09}), the mass-size relation (e.g. \citealt{williams10}), the Tully-Fisher relation (e.g. \citealt{mocz12}) and the fundamental plane along with studies of galaxy formation, evolution and internal structure (e.g. \citealt{gunawardhana12}). In the current data-rich era it is typically the systematic rather than random errors which now dominate this process. The drive to reduce systematic errors and further constrain models therefore demands high-precision galaxy photometry and high-fidelity spatial profiling, hence the need for more sensitive imagery with greater spatial resolution over substantive areas of sky. To this extent, a number of wide-field surveys such as the Sloan Digital Sky Survey (SDSS; \citealt{york00,abazajian08}), the 2-Micron All Sky Survey (2MASS; \citealt{skrutskie06}), the Panoramic Survey Telescope and Rapid Response System (Pan-STARRS), the \textit{Galaxy Evolution Explorer} (\textit{GALEX}; \citealt{martin05}) Medium Imaging Survey (MIS; \citealt{morrissey07}), the VLT Survey Telescope (VST) Kilo-Degree Survey (KIDS; \citealt{arnaboldi07}), the UKIRT (UK InfraRed Telescope) Infrared Deep Sky Survey (UKIDSS; \citealt{lawrence07}) and the VISTA (Visible and Infrared Survey Telescope for Astronomy) Kilo-degree INfrared Galaxy survey (VIKING; \citealt{arnaboldi07}) have been initiated with the aim of probing large areas of the sky deeper than preceeding surveys at a variety of wavelengths. In particular the advances in the near infrared (NIR) have been rapid and substantial, progressing from 2MASS to UKIDSS and currently VISTA VIKING.

Measuring the total flux from a galaxy, particularly in the NIR, is not a trivial undertaking --- beyond a certain radius, the surface brightness drops sufficiently to be indistinguishable from background noise. Simple but crude aperture photometry methods such as setting a fixed aperture size or detection threshold ignores this problem completely. Other, more complex methods, such as those suggested by \citet{kron80} or \citet{petrosian76}, are subject to bias with galaxy morphology \citep{graham05b}. An alternative approach is to fit a light profile and integrate to infinity or some fixed number of effective radii (e.g. \citealt{kelvin12}). One general and commonly used profile is the S\'{e}rsic profile \citep{sersic63,sersic68,graham05}. The S\'{e}rsic equation describes how the intensity of light $I$ from a galaxy varies as a function of radius $r$ as follows:

\begin{equation} \label{eq:sersic}
I(r) = I_e \exp\left[-b_n\left(\left(\frac{r}{r_e}\right)^{1/n} - 1\right)\right]
\end{equation}

where $I_e$ is the intensity at the effective radius $r_e$ which encloses half of the total light and $n$ is the S\'{e}rsic index. $b_n$ is such that $\Gamma(2n) = 2\gamma(2n, b_n)$, where $\Gamma$ and $\gamma$ are the complete and incomplete gamma functions respectively \citep{ciotti91}. Notable S\'{e}rsic indices include the Gaussian $n = 0.5$ profile, the exponential $n = 1$ profile associated with galactic disks \citep{freeman70} and the \citet{dev48} $n = 4$ profile initially proposed for elliptical galaxies.\footnote{For a comparison of Kron, Petrosian and S\'{e}rsic magnitudes and how they vary with central concentration, see \citet{hill11}.} 

S\'{e}rsic parameters, magnitudes and derived quantities can then be assembled as a function of distance and/or environment and compared to similarly defined samples drawn from numerical simulations. The S\'{e}rsic index and the bulge luminosity are particularly interesting measurements as they are known to be correlated with the mass of the central supermassive black hole (e.g. \citealt{graham07,novak06, vika12, savorgnan13}). 

\begin{sloppypar}
The Galaxy and Mass Assembly (GAMA) project \citep{driver11} is a spectroscopic and multi-wave\-length imaging survey of galaxies aimed at exploring the existence, evolution and spatial extent of mass and energy distributions from kpc to Mpc scales. GAMA combines redshifts from the AAOmega spectograph at the 3.9~m Anglo-Australian Telescope, optical imaging from SDSS and NIR imaging from the Large Area Survey component of UKIDSS (UKIDSS-LAS) in three 12 (RA) $\times$ 5 (Dec) degree equatorial regions centered around $09^h$, $12^h$ and $14.5^h$ and (G09, G12 and G15 respectively). In the future, the GAMA team will ingest data from \textit{GALEX} MIS (UV), VST KIDS (optical), VISTA VIKING (NIR), \textit{WISE} (mid infrared), \textit{Herschel}--ATLAS (far infrared), ASKAP DINGO (21 cm) and GMRT (325 MHz) and observe additional regions in the southern sky. In this paper, we use the GAMA dataset and processing pipeline to establish $K_s$-band limits to which various photometric and structural measurements, made using 2MASS, UKIDSS-LAS and VISTA VIKING data, \textbf{are reliable}.
\end{sloppypar}

The structure of this paper is as follows. We give a brief overview of the leading wide-area NIR surveys --- 2MASS, UKIDSS-LAS and VISTA VIKING --- in Section~\ref{sec:data}. In Section~\ref{sec:processing}, we describe how we standardised the data, assembled mosaics, estimated galaxy number counts and recovered S\'{e}rsic profile fits and photometry for a sample of 50,123 GAMA galaxies. In Section~\ref{sec:results}, we use these to derive qualitative and quantitative limits to which various photometric and structural parameters \textbf{are reliable}. This data should be useful for the design of future surveys based on these data. Magnitudes are given in the AB system unless stated otherwise. 

\section{DATA}
\label{sec:data}

\subsection{2MASS}

\begin{sloppypar}
During the period June 1997 -- February 2001, the 2MASS project \citep{skrutskie06} surveyed 99.998\% of the sky in the NIR. Each area of sky was viewed for 7.8~s in the $JHK_s$ bands using two $1.3$~m telescopes located at Whipple Observatory, USA and Cerro Tololo, Chile. Each telescope was equipped with three $256 \times 256$ pixel Rockwell CCDs with a pixel scale of 2 arcsec per pixel. The three bands were imaged simulaneously. The survey achieved a limiting (AB) magnitude of $J = 16.7$~mag, $H = 16.5$~mag and $K_s = 16.2$~mag at $10 \sigma$, $1 \sigma$ photometric uncertainty of $< 0.03$ mag, an astrometric accuracy of $\sim 0.1$~arcsec and typical seeing full width at half maximum (FWHM) of 2.5 -- 3~arcsec. In addition to images covering most of the NIR sky, the survey also produced a 471 million object point source catalog and an extended source catalog of 1.6~million objects. The 2MASS All-Sky Data Release, being the entire 2MASS dataset,  was released to the public in 2003 March\footnote{The data are accessible at \newline \url{http://www.ipac.caltech.edu/2mass/releases/allsky/}.}. 2MASS has been responsible for over 1000 publications, with an aggregate of more than 32,000 citations \citep{driver13}.
\end{sloppypar}

\subsection{UKIDSS-LAS}

\begin{sloppypar}
UKIDSS was a seven year, five survey programme which commenced in May 2005 utilizing the Wide Field CAMera (WFCAM) on the 3.8 m UKIRT atop Mauna Kea, Hawaii \citep{lawrence07}. UKIDSS-LAS aimed to provide a NIR complement to SDSS, overlapping stripes 9--16, 26--33 and part of stripe 82. UKIDSS-LAS covered 4028 deg$^2$ of sky in the\textit{YJHK$_s$} bands to limiting magnitudes of $Y = 20.9$~mag, $J = 20.8$~mag (after two passes), $H = 20.0$~mag and $K_s = 20.1$~mag at $5 \sigma$. UKIDSS-LAS achieved a typical seeing FWHM of $< 1.2$~arcsec, a photometric uncertainty of $< 0.02$~mag and an astrometric accuracy of 0.1~arcsec.
\end{sloppypar}

WFCAM \citep{casali07} is an array of four $2048 \times 2048$~pixel Rockwell CCDs with a pixel scale of 0.4~arcsec per pixel.  Raw images, or `pawprints', are compressed and stored using one FITS binary HDU per CCD. Four pawprints are stitched together to form a contiguous image called a `tile', covering a 0.78 deg$^2$ field of view (including overlaps) with a combined exposure time of 40~s. UKIDSS calibration is tied at bright magnitudes to 2MASS (see \citealt{hodgekin09}).

\subsection{VISTA VIKING}

The VISTA VIKING survey \citep{arnaboldi07}, commenced in April 2010 and utilizes the VISTA InfraRed Camera (VIRCAM; \citealt{dalton06}) on the 4m VISTA telescope, located at Paranal, Chile, and operated by ESO. VIKING will image 1500 deg$^2$ of sky in the $ZYJHK_s$ bands to target limiting magnitudes of $Z = 23.1$~mag, $Y = 22.3$~mag, $J = 22.1$~mag, $H = 21.5$~mag, and $K_s = 21.2$~mag at $5 \sigma$. VIKING is targetting a seeing FWHM of $< 1.0$~arcsec.

VIRCAM is a $4 \times 4$ array of $2048 \times 2048$ pixel Ray\-theon CCDs with a pixel scale of 0.339 arcsec per pixel. Six pawprints, exposed for 10~s each, are assembled into a VISTA tile, which has a field of view 1.65~deg in diameter. Most of the sky in a given tile is covered at least twice. 

Both UKIDSS-LAS and VIKING data are processed using the Vista Data Flow System \citep{emerson04}, with final data products being stored in the WFCAM Science Archive (WSA)\footnote{\url{http://surveys.roe.ac.uk/wsa/}} and the VISTA Science Archive (VSA)\footnote{\url{http://horus.roe.ac.uk/vsa/}} respectively \citep{hambly08}. VISTA calibration is also tied at bright magnitudes to 2MASS.

\section{IMAGE ACQUISITION AND PROCESSING}
\label{sec:processing}

\begin{figure}
\begin{center}
\includegraphics[width=3in]{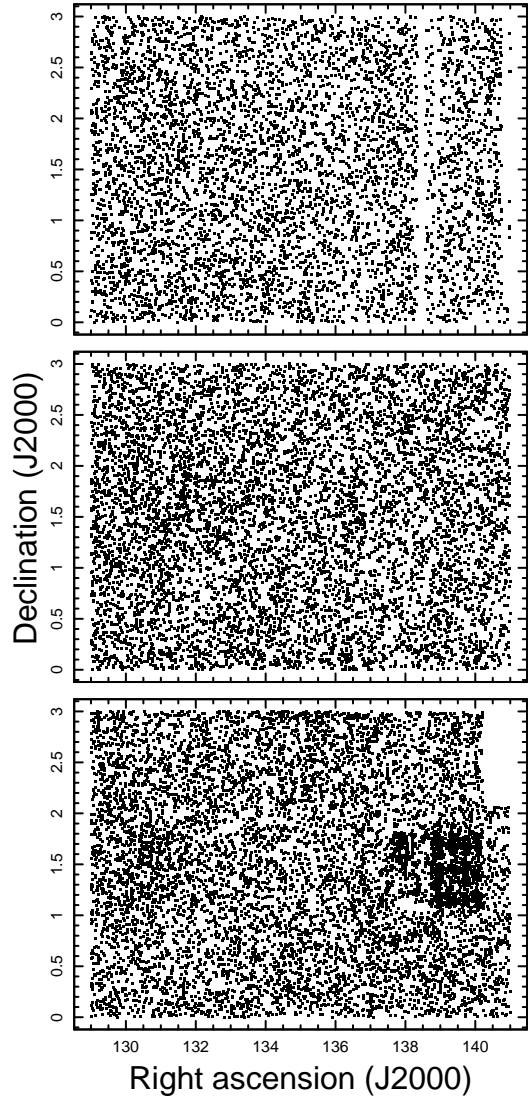}
\caption{[New figure]. Coverage of the G09 region above 0\degree~declination in the three surveys -- top: 2MASS, center: UKIDSS-LAS, bottom: VISTA VIKING.}
\label{fig:coverage}
\end{center}
\end{figure}

\begin{sloppypar}
\textbf{For this work, we use a subset of the G09 region above 0\degree~declination. This region, and the coverage of the surveys within it are illustrated in Figure \ref{fig:coverage}.} We downloaded 1500 2MASS tiles from the 2MASS All-Sky Data Release and 315 early release VIKING pawprints from VSA containing $K_s$ band imagery of G09 (centered on 09$^h$00$^m$30$^s$.0 $+$00\degree15'00''.0). 928 UKIDSS-LAS images from the fourth data release were obtained from WSA and processed by \citet{hill11}.  We did not use VISTA VIKING tile images from VSA because they were constructed without background subtraction and therefore exhibit significant systematic large scale sky variations. 

\end{sloppypar}

To process the VISTA VIKING and 2MASS images into single large-area mosaics, we applied a procedure similar to that detailed in \citet{hill11}, which we summarize below.

\begin{sloppypar}
We use a custom pipeline to extract the exposure time (t), airmass ($a = \sec \chi_\mathrm{mean} - 1 = \mathrm{(AM_{START} + AM_{END})}/2$) and extinction (ext) from the FITS header for each image HDU. We determine an AB magnitude zero point for VISTA VIKING images using the equation: 
\end{sloppypar}

\begin{equation} \label{eq:norm}
\mathrm{ZP_{total}} = \mathrm{ZP_{mag}} + 2.5 \log(t) - \mathrm{ext} \times a + \mathrm{ABV_K}
\end{equation}
\\
where $ABV_K = 1.9$~mag is the AB magnitude of Vega in the $K_s$ band \citep{hewitt06}. The 2MASS data set is already calibrated for exposure time; in this case the AB zero point is given by:

\begin{equation} \label{eq:norm2}
\mathrm{ZP_{total}} = \mathrm{ZP_{mag}} - \mathrm{ext} \times a + \mathrm{ABV_K}
\end{equation}

\begin{sloppypar}
We convert each image to a common zero point of 1 ADU = 30~mag in 1 second by multiplying each pixel by a constant $10^{-0.4(\mathrm{ZP_{total}} - 30)}$. The sky noise, sky background, read noise and saturation threshhold were scaled by the same factor and the gain was adjusted to keep the number of electrons unaltered. 
\end{sloppypar}

We assembled mosaics of the renormalized images using the \textsc{swarp} (v2.19.1) utility \citep{bertin02}. \textsc{swarp} is a multithread-capable image stitching, \textbf{stacking and warping tool. The construction of mosaics} minimizes the effect of objects being split across frames and differing zero points between frames. As we intend to match objects between surveys using astrometric measurement, it is imperative we create perfectly matched mosaics. We resample the images using the LANCZOS3 algorithm to a pixel scale of $0.339 \times 0.339$ arcsec in the TAN projection system \citep{calabretta02}, generating a $204500 \times 79700$ pixel image centered around 09$^h$00$^m$30s.0 +00\degree15'00''.0. We set \textsc{swarp} to subtract the background using a mesh of $512 \times 512$~pixels ($1024 \times 1024$~arcsec, $205 \times 205$~arcsec and $174 \times 174$~arcsec for 2MASS, UKIDSS-LAS and VISTA VIKING data respectively) and a background filter size of $3 \times 3$ meshes. The mosaics are $\sim60$~GB in size.

\subsection{Qualitative inspection}

\begin{figure*}
\begin{center}
\includegraphics[width=0.9in]{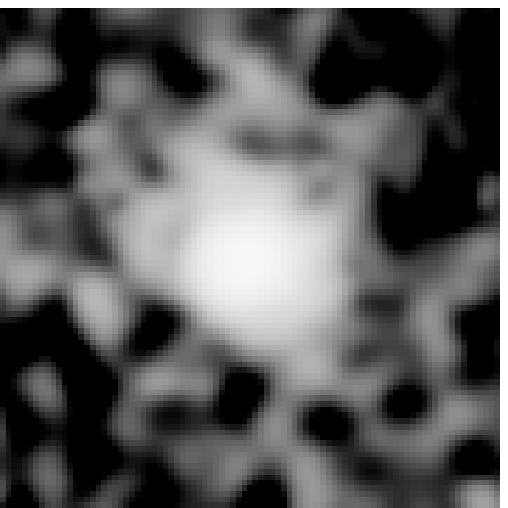}
\includegraphics[width=0.9in]{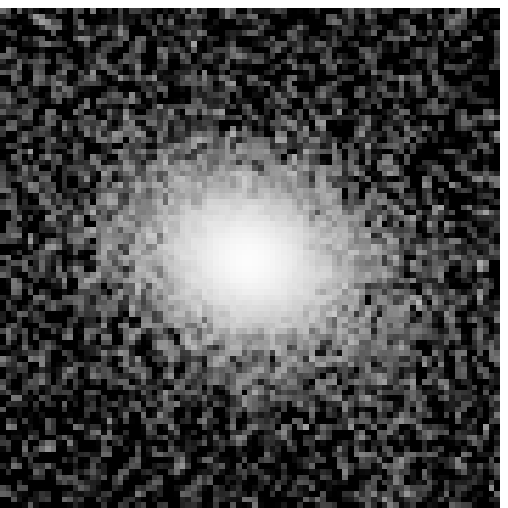}
\includegraphics[width=0.9in]{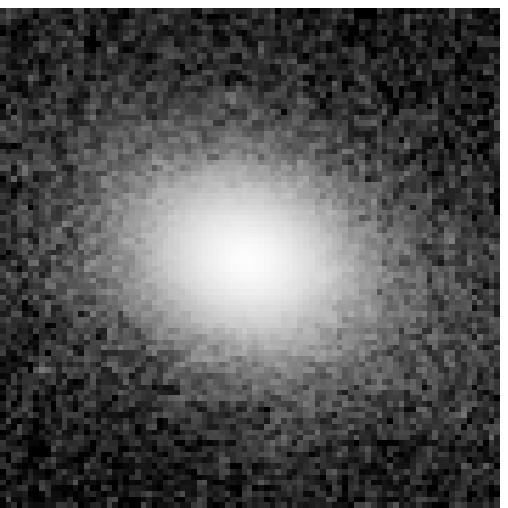}
\hspace{0.8cm}
\includegraphics[width=0.9in]{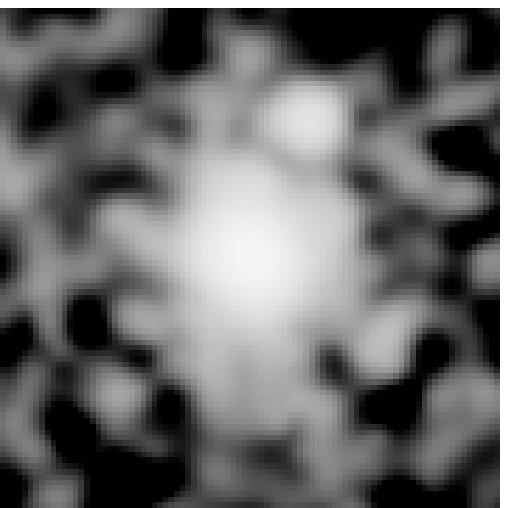}
\includegraphics[width=0.9in]{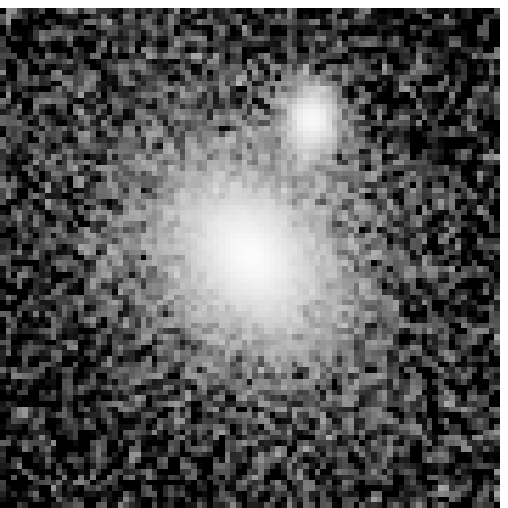}
\includegraphics[width=0.9in]{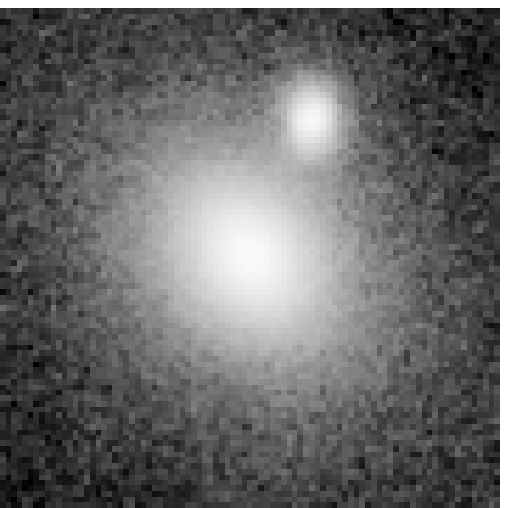}
\\
\includegraphics[width=0.9in]{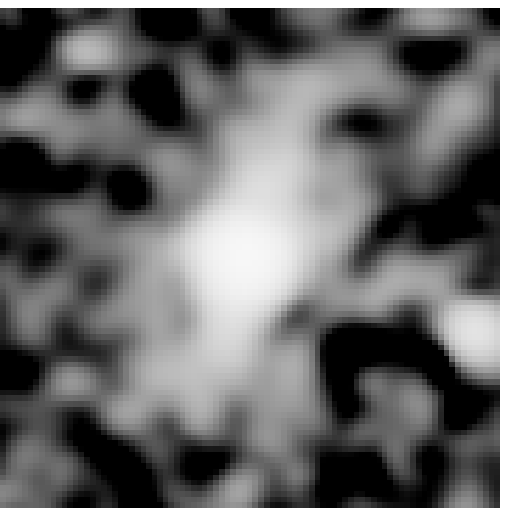}
\includegraphics[width=0.9in]{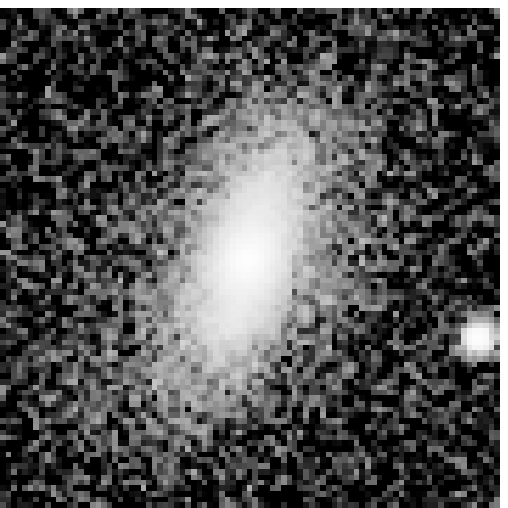}
\includegraphics[width=0.9in]{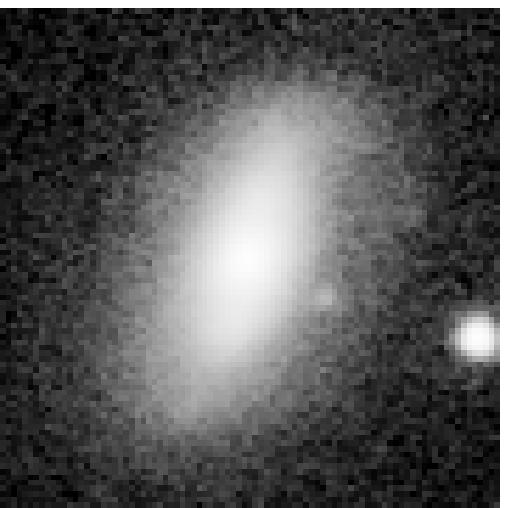}
\hspace{0.8cm}
\includegraphics[width=0.9in]{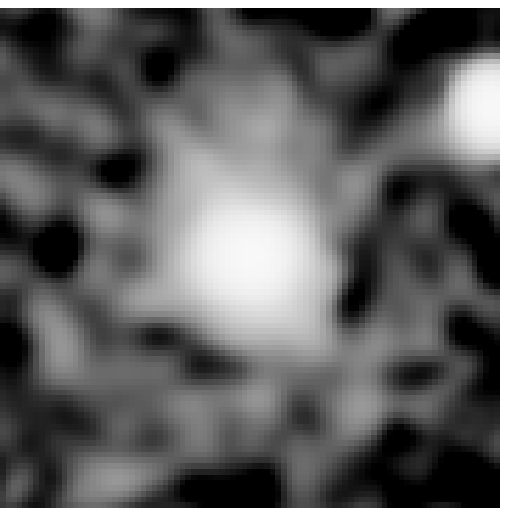}
\includegraphics[width=0.9in]{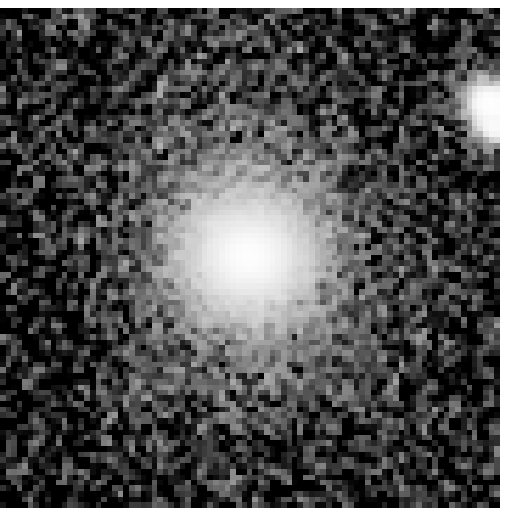}
\includegraphics[width=0.9in]{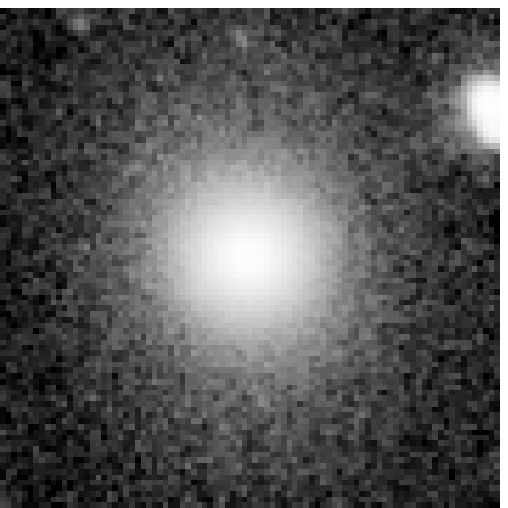}
\\
\includegraphics[width=0.9in]{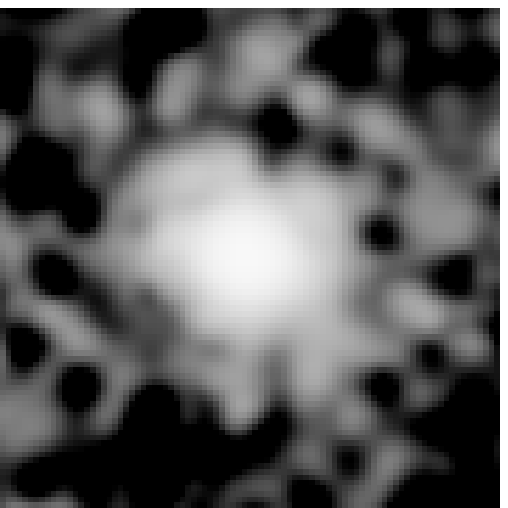}
\includegraphics[width=0.9in]{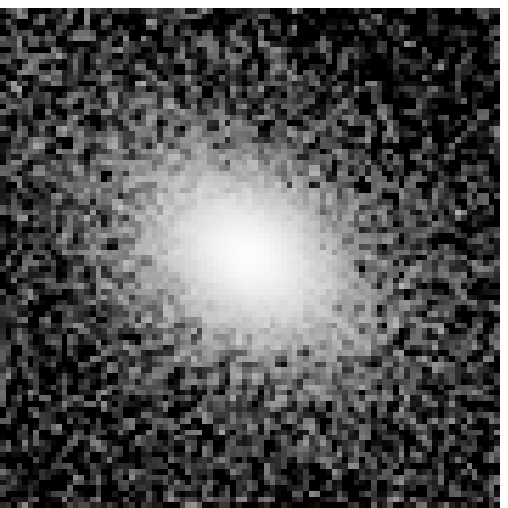}
\includegraphics[width=0.9in]{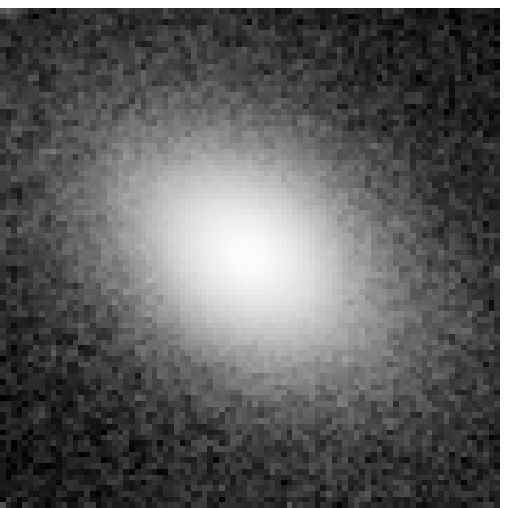}
\hspace{0.8cm}
\includegraphics[width=0.9in]{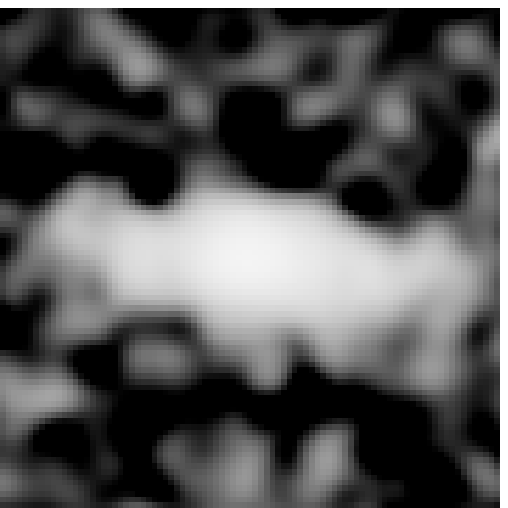}
\includegraphics[width=0.9in]{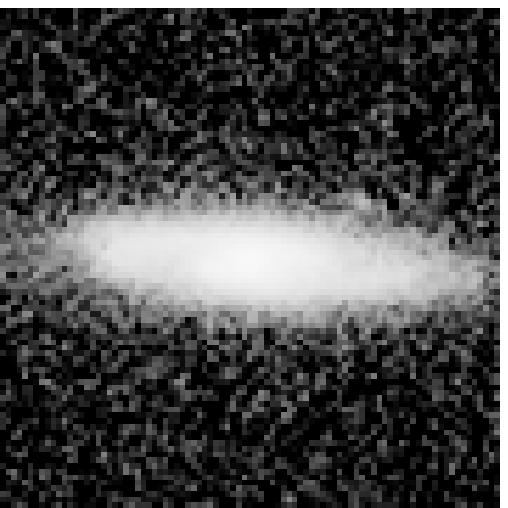}
\includegraphics[width=0.9in]{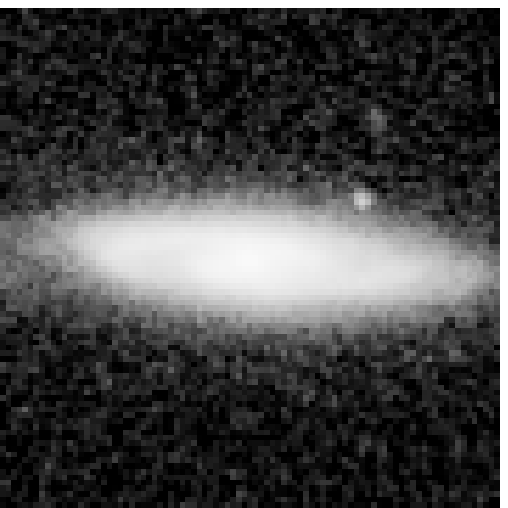}
\\
\includegraphics[width=0.9in]{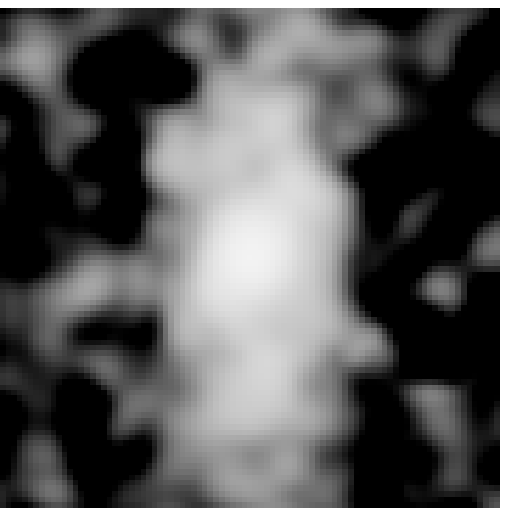}
\includegraphics[width=0.9in]{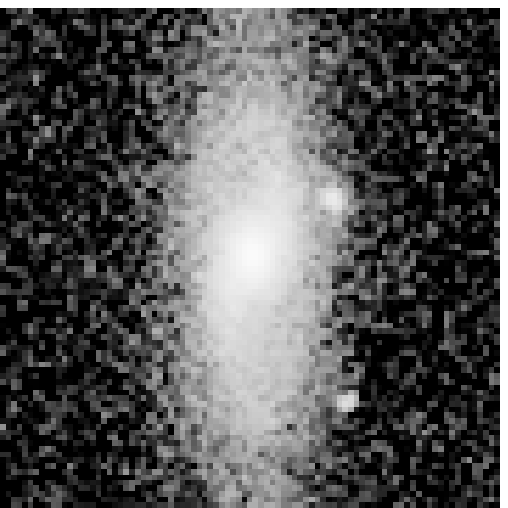}
\includegraphics[width=0.9in]{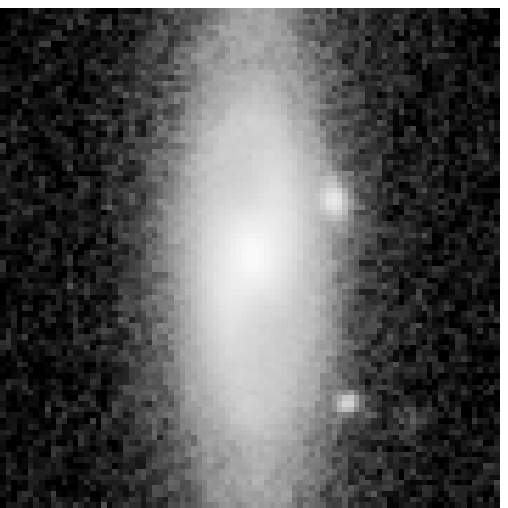}
\hspace{0.8cm}
\includegraphics[width=0.9in]{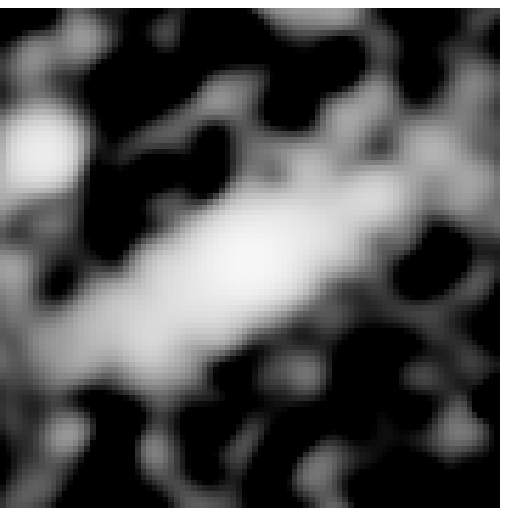}
\includegraphics[width=0.9in]{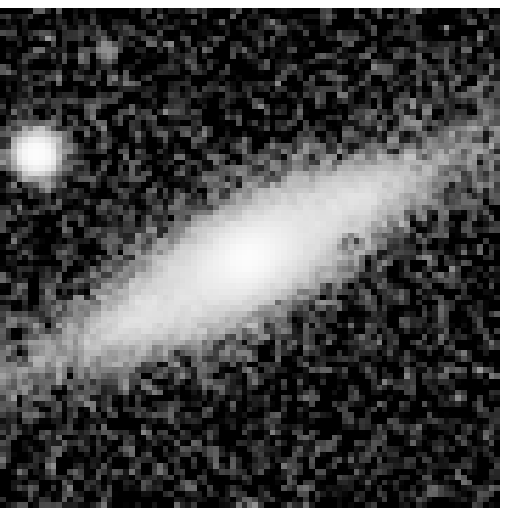}
\includegraphics[width=0.9in]{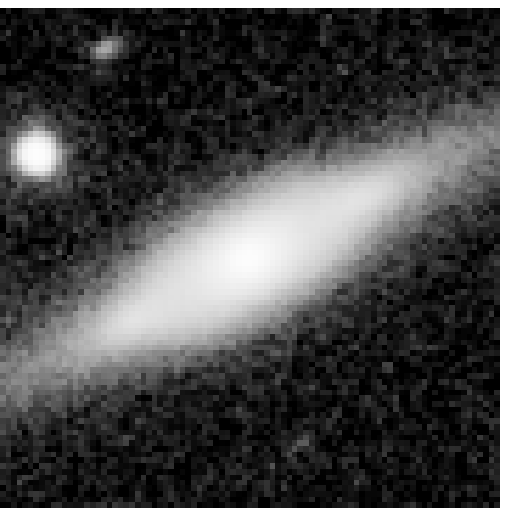}
\\
\includegraphics[width=0.9in]{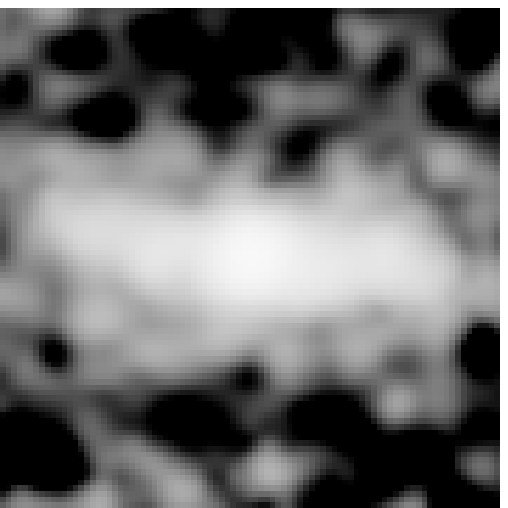}
\includegraphics[width=0.9in]{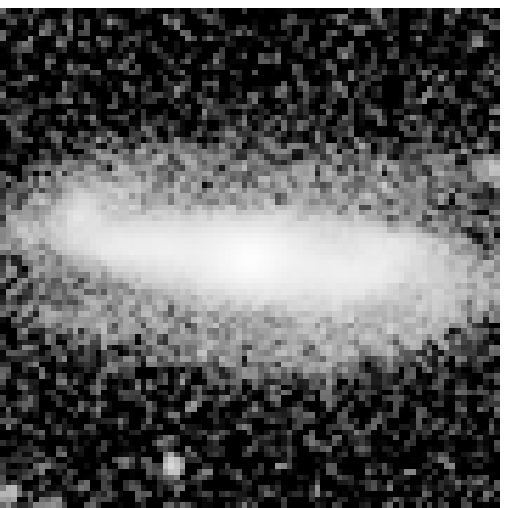}
\includegraphics[width=0.9in]{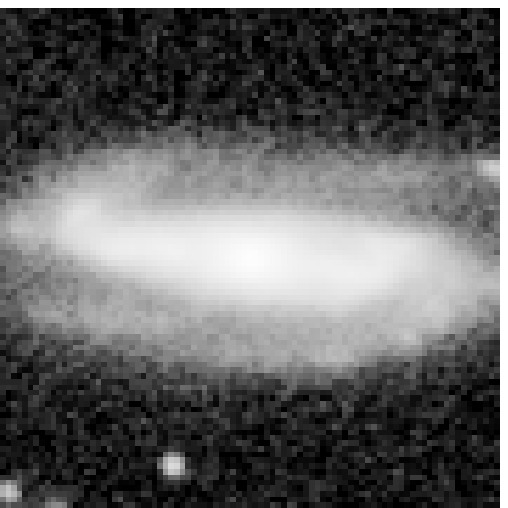}
\hspace{0.8cm}
\includegraphics[width=0.9in]{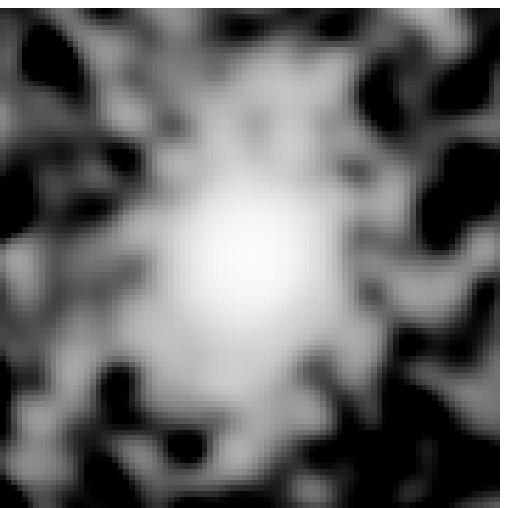}
\includegraphics[width=0.9in]{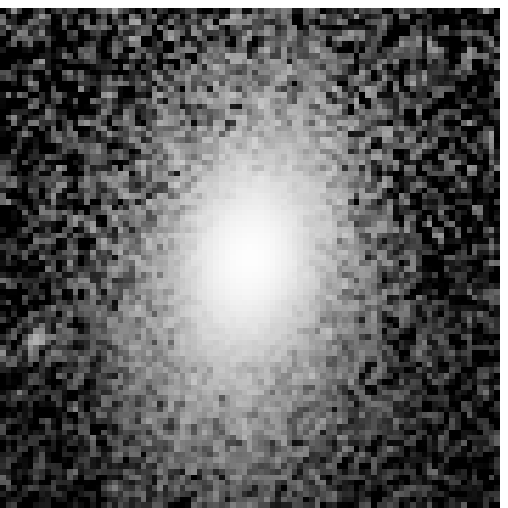}
\includegraphics[width=0.9in]{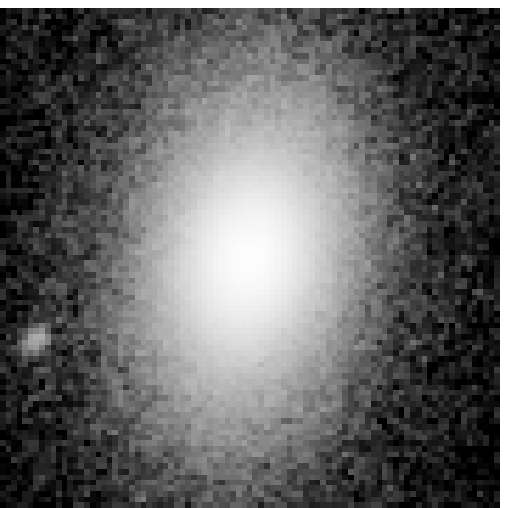}
\\
\includegraphics[width=0.9in]{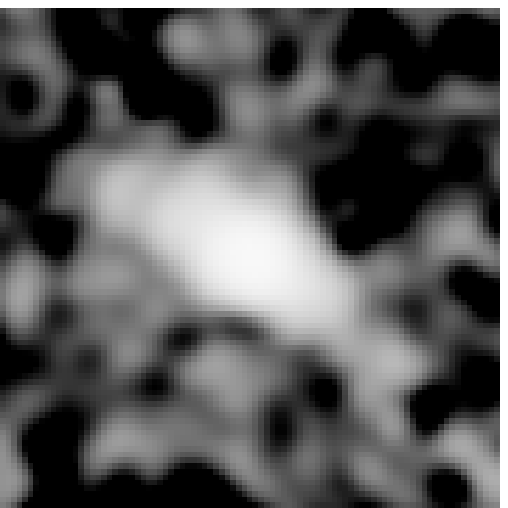}
\includegraphics[width=0.9in]{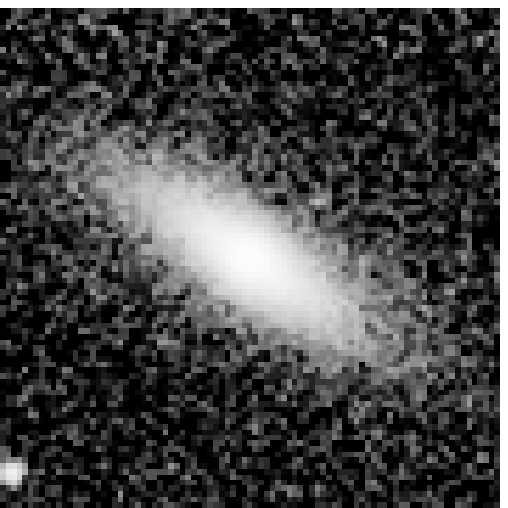}
\includegraphics[width=0.9in]{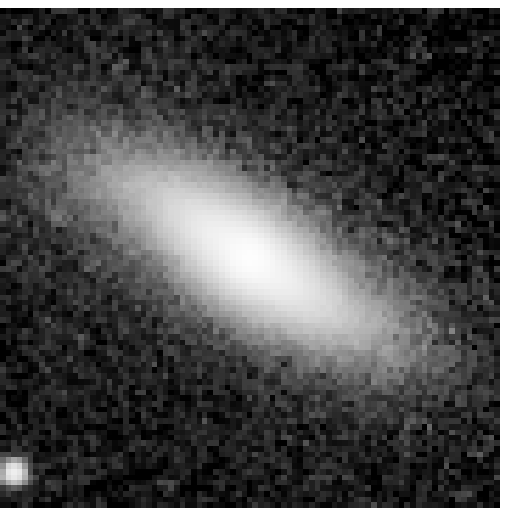}
\hspace{0.8cm}
\includegraphics[width=0.9in]{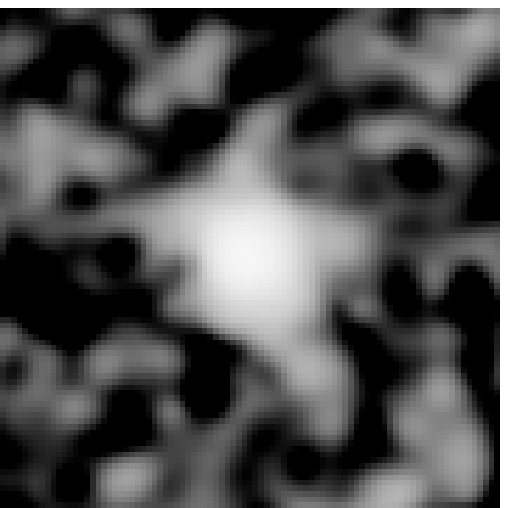}
\includegraphics[width=0.9in]{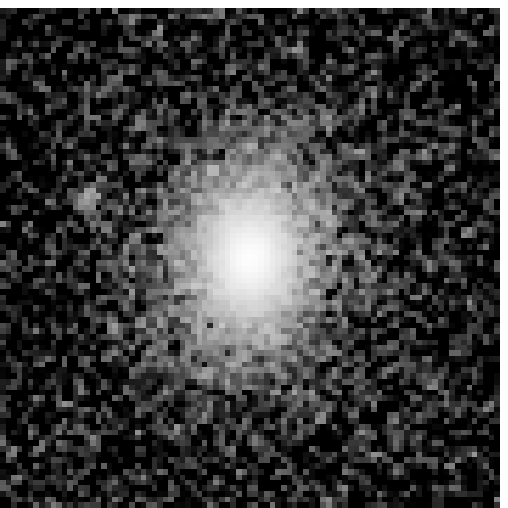}
\includegraphics[width=0.9in]{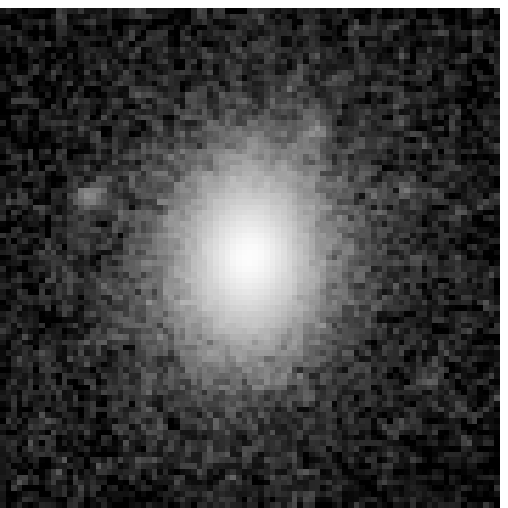}
\\
\includegraphics[width=0.9in]{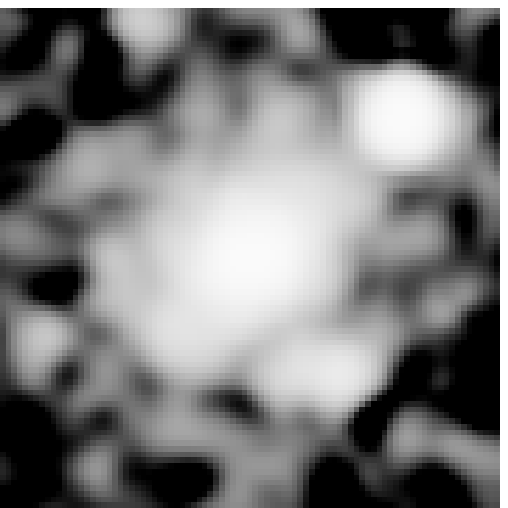}
\includegraphics[width=0.9in]{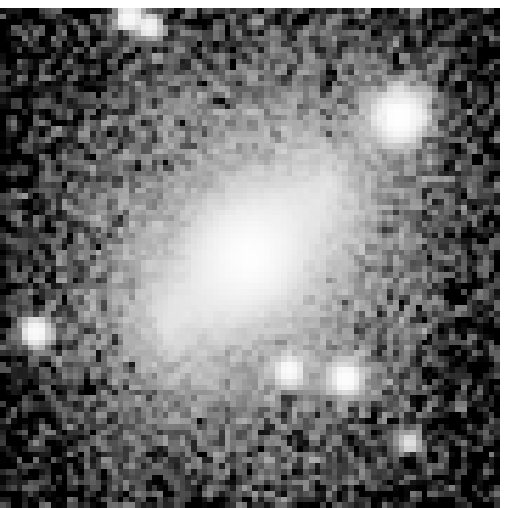}
\includegraphics[width=0.9in]{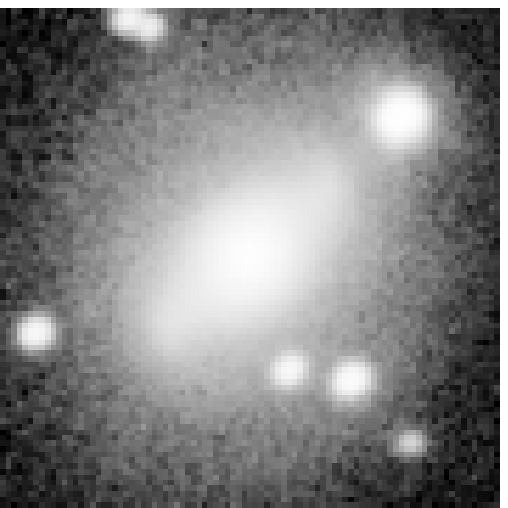}
\hspace{0.8cm}
\includegraphics[width=0.9in]{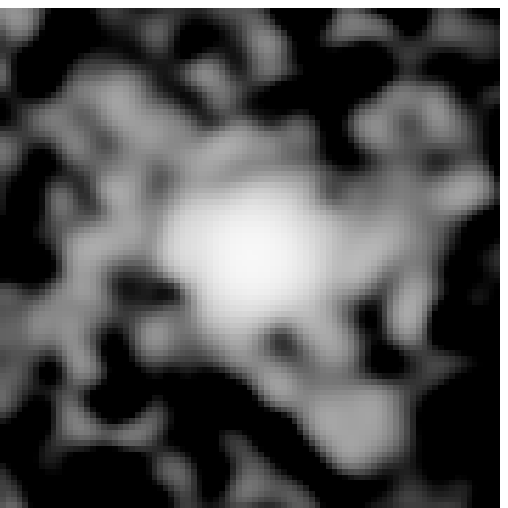}
\includegraphics[width=0.9in]{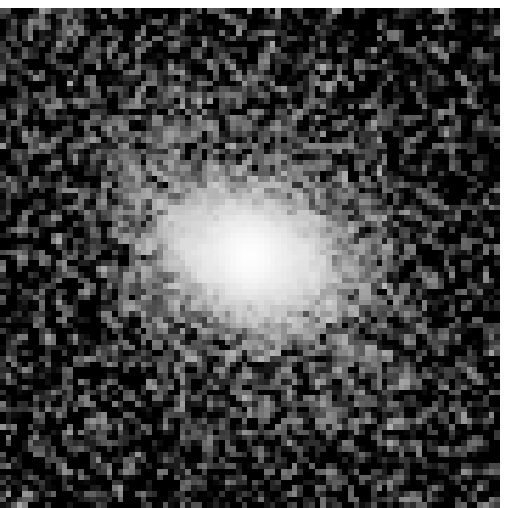}
\includegraphics[width=0.9in]{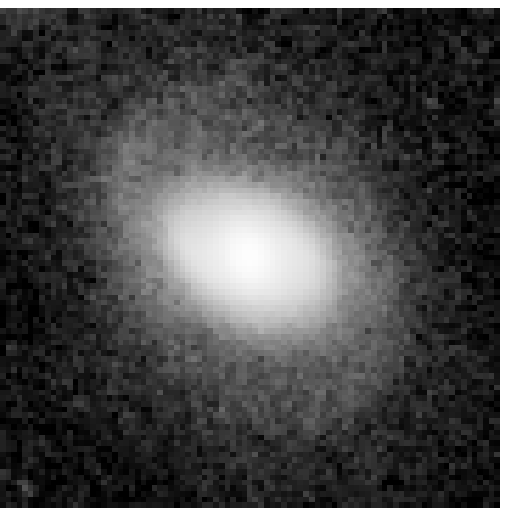}
\\
\includegraphics[width=0.9in]{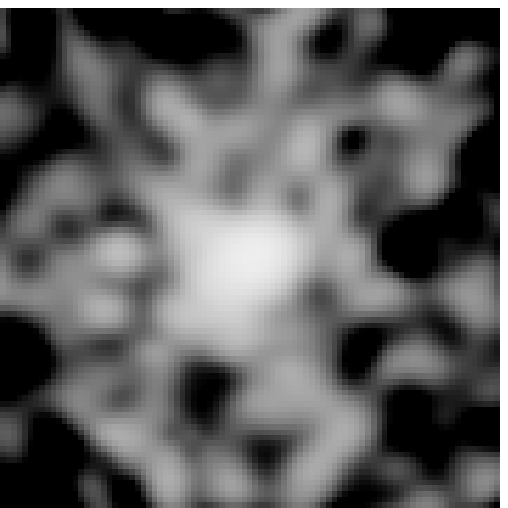}
\includegraphics[width=0.9in]{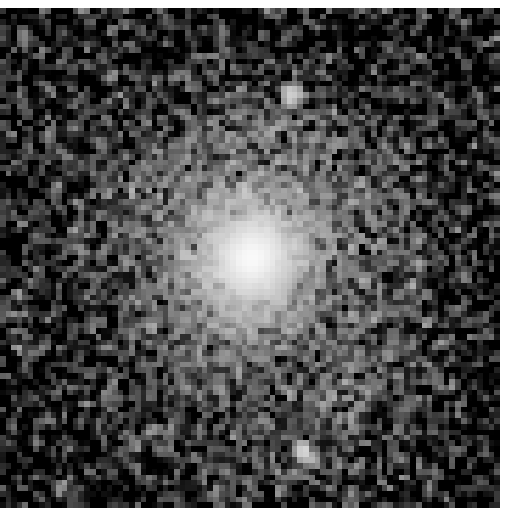}
\includegraphics[width=0.9in]{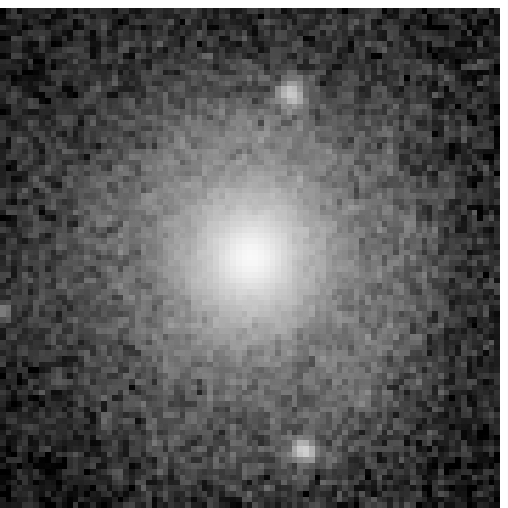}
\hspace{0.8cm}
\includegraphics[width=0.9in]{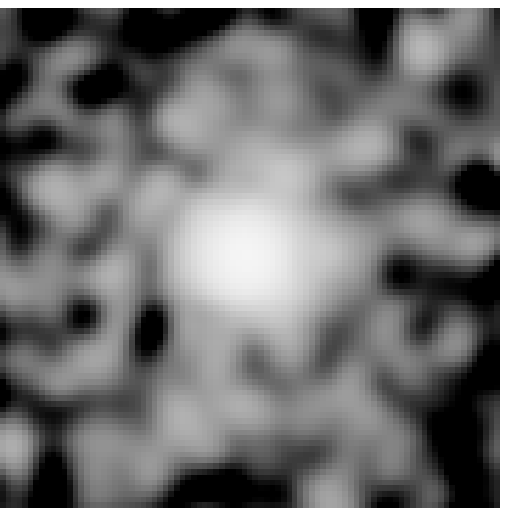}
\includegraphics[width=0.9in]{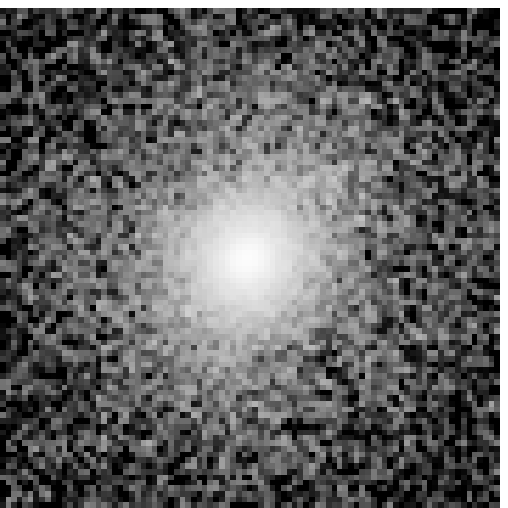}
\includegraphics[width=0.9in]{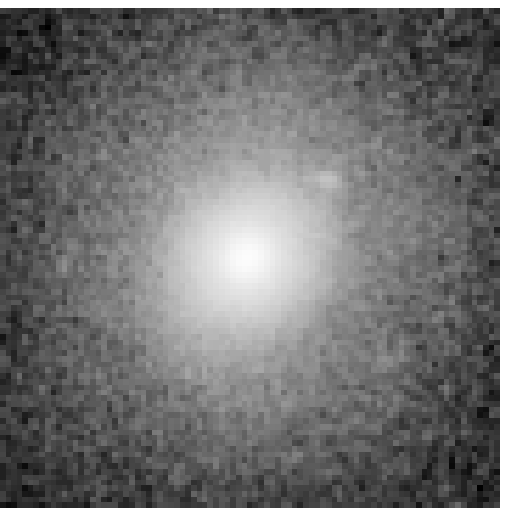}
\\
\includegraphics[width=0.9in]{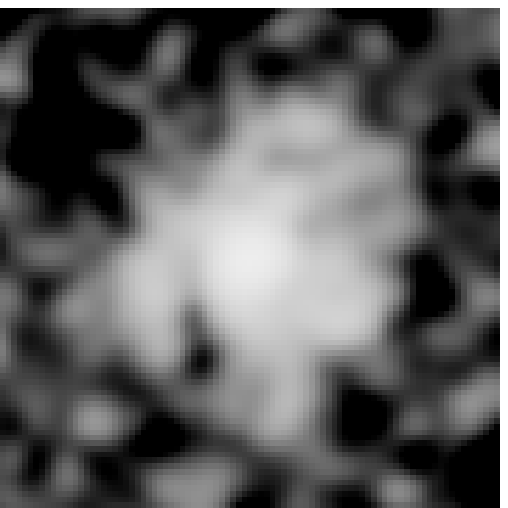}
\includegraphics[width=0.9in]{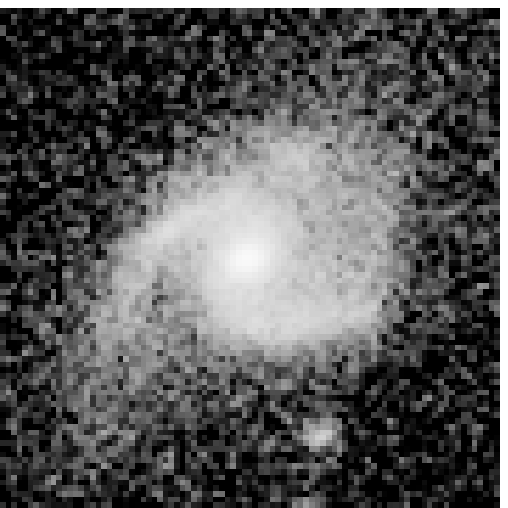}
\includegraphics[width=0.9in]{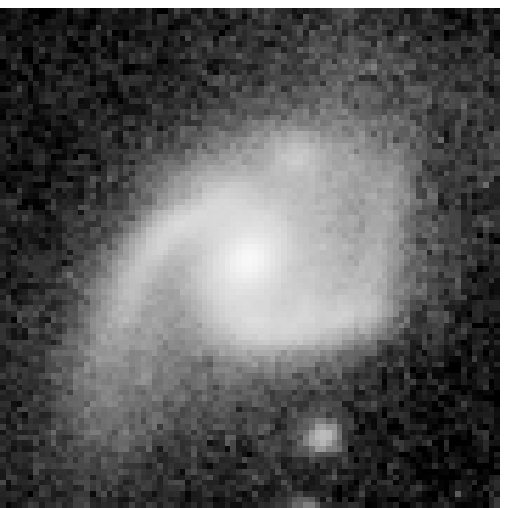}
\hspace{0.8cm}
\includegraphics[width=0.9in]{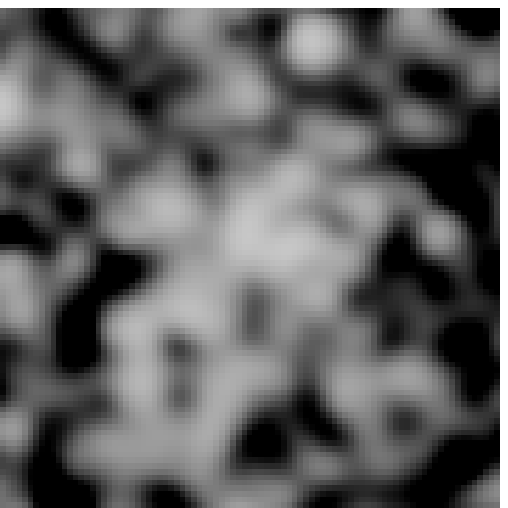}
\includegraphics[width=0.9in]{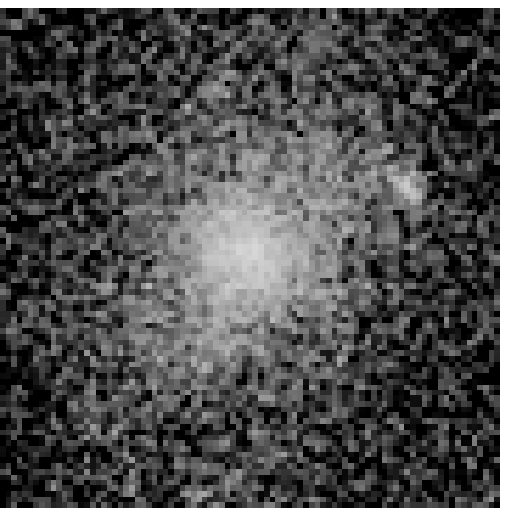}
\includegraphics[width=0.9in]{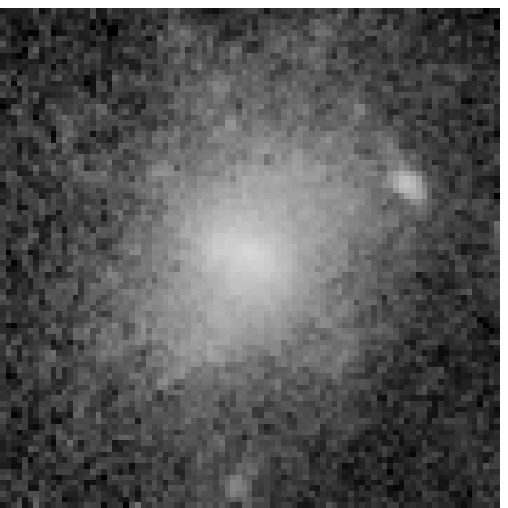}

\caption{$K_s$ band images of a sample of G09 galaxies; left: 2MASS, center: UKIDSS-LAS, right: VISTA VIKING; from left to right then top to bottom: GAMA IDs 204839, 205085, 209698, 214250, 214363, 216670, 278390, 278802, 278847, 279905, 279908, 303010, 303096, 325007, 3768802, 380578, 526797 and 600168. The cutout radius is 15 arcsec.}
\label{fig:pictures}
\end{center}
\end{figure*}
\begin{sloppypar}
Figure \ref{fig:pictures} shows $K_s$ band cutouts from these mosaics for a selection of G09 galaxies with $r_{\mathrm{Petrosian}} < 15$~mag (as defined by the GAMA database; \citealt{driver11}). Figure \ref{fig:colorpictures} presents cutouts of the first four galaxies in Figure \ref{fig:pictures} combined with corresponding cutouts from existing resampled GAMA SDSS DR6 mosaics \citep{hill11} to produce $K_sig$ (=RGB) false colour images. The signal to noise ratio (S/N) and spatial resolution of the VISTA VIKING data are comparable to SDSS and are a marked improvement over UKIDSS-LAS, which in turn improves over 2MASS. The non-detection of the bottom (GAMA ID 600618) galaxy in the 2MASS dataset underscores the importance increased depth plays in the detection of low surface brightness (LSB) galaxies and galaxy components such as disks. Indeed, \citet{bell03} observe that non-detections of disk-dominated systems are a widespread phenomenon in the 2MASS data set. The increase in depth between UKIDSS-LAS and VISTA VIKING is also clear with the non-detection of the disks of some galaxies and the increased visual size of the ellipticals despite identical scaling.

\begin{figure}
\begin{center}
\includegraphics[width=0.9in]{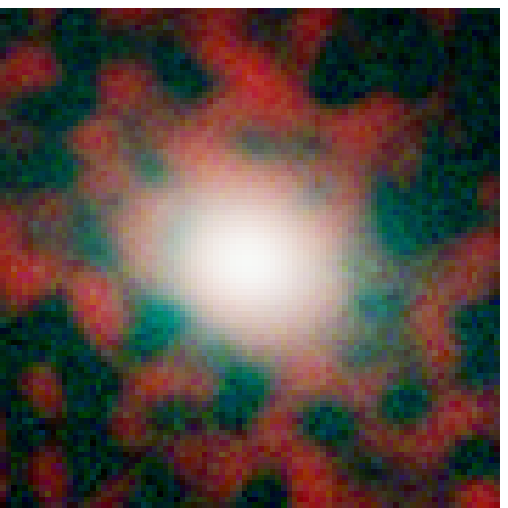}
\includegraphics[width=0.9in]{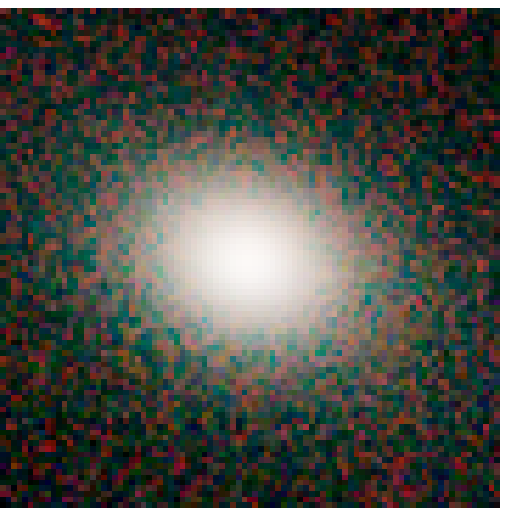}
\includegraphics[width=0.9in]{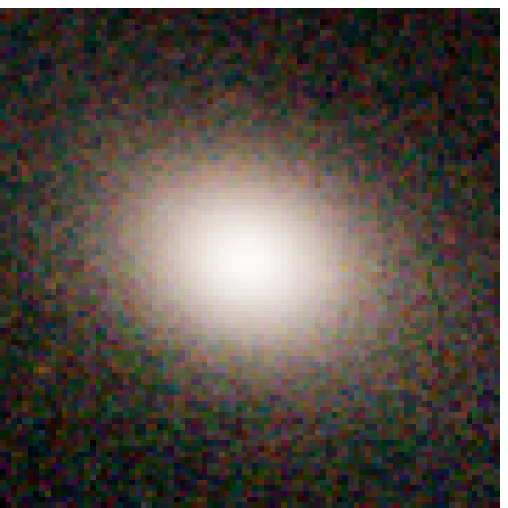}
\\
\includegraphics[width=0.9in]{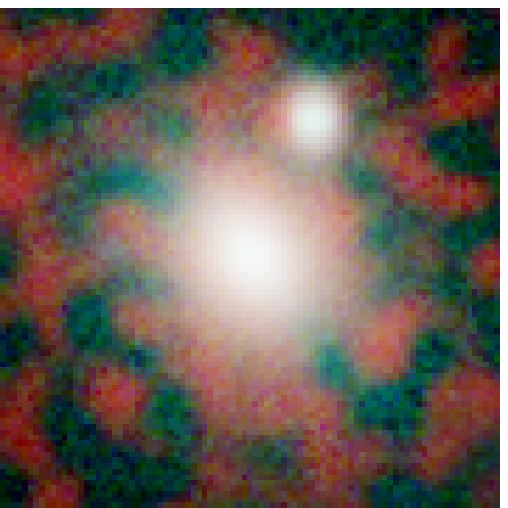}
\includegraphics[width=0.9in]{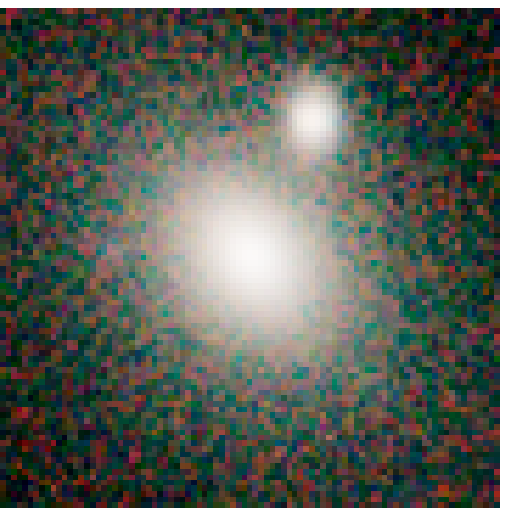}
\includegraphics[width=0.9in]{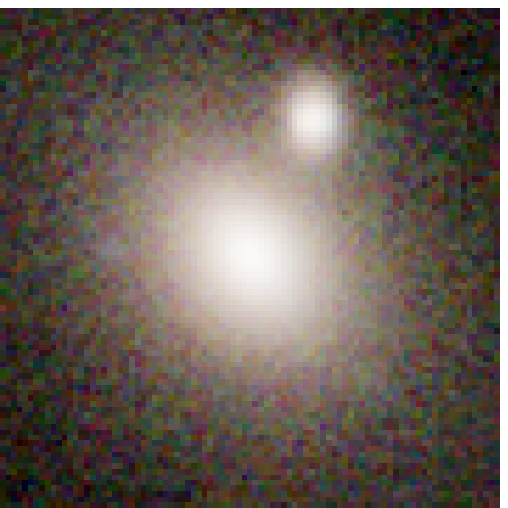}
\\
\includegraphics[width=0.9in]{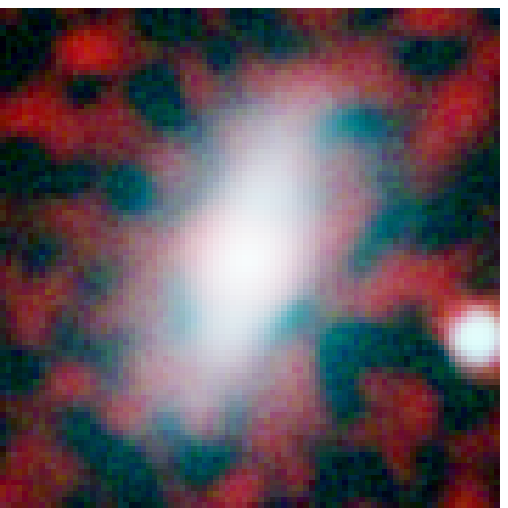}
\includegraphics[width=0.9in]{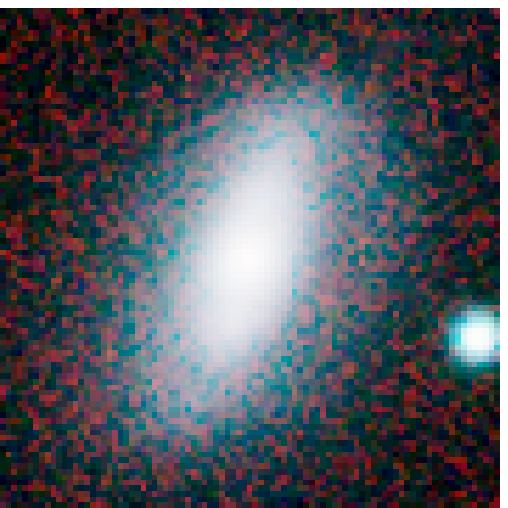}
\includegraphics[width=0.9in]{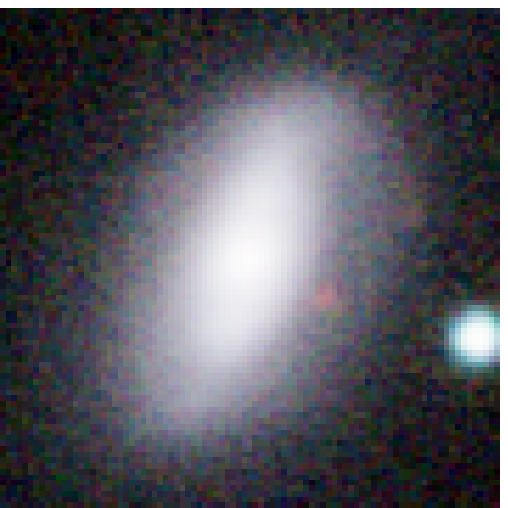}
\\
\includegraphics[width=0.9in]{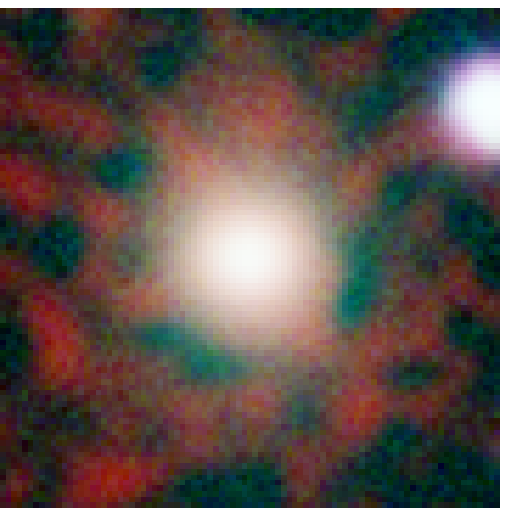}
\includegraphics[width=0.9in]{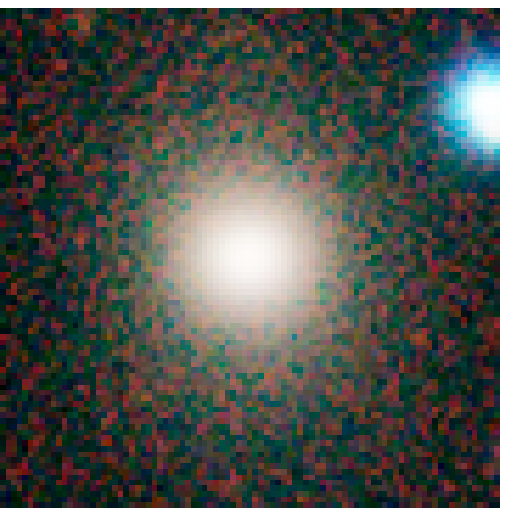}
\includegraphics[width=0.9in]{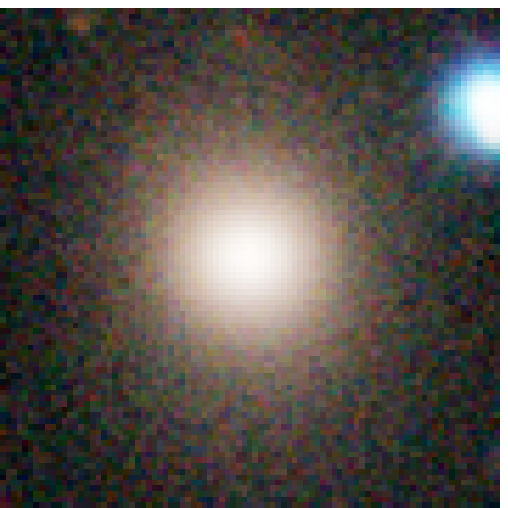}

\caption{$K_sig$ = RGB false colour images of the first four galaxies (GAMA IDs 204839, 205085, 209698 and 214250)  in Figure \ref{fig:pictures}; left: 2MASS, center: UKIDSS-LAS, right: VISTA VIKING.}
\label{fig:colorpictures}
\end{center}
\end{figure}

\end{sloppypar}
To estimate the randomness of background noise, we visually selected 11 empty $7 \times 7$ pixel regions from the mosaics and computed the standard deviation of the mean pixel value of each region, which we denote as $q$. We compared this against a similar calculation after randomly swapping pixels between regions. A $q/q_{swapped}$ of 1 is indicative of perfectly uncorrelated noise while $q/q_{swapped} >> 1$ alludes to non-pixel scale systematic variations. 2MASS images show a high correlation between adjacent pixels ($q/q_{swapped} \sim 7$) due to significant upsampling in the mosaic creation process; this is readily apparent in Figures \ref{fig:pictures} and \ref{fig:colorpictures}. Surprisingly, VISTA VIKING ($q/q_{swapped} \sim 2.7$) exhibits higher background correlation than UKIDSS-LAS ($q/q_{swapped} \sim 1$) with $q_\mathrm{VIKING} < q_\mathrm{UKIDSS}$ by about 10\% prior to the swap. This appears to be the result of vastly lower random noise levels in the VISTA VIKING data --- systematic variations \textbf{in noise intrinsic to the VIRCAM detector} on the $\sim10$~pixel scale have become important. These variations are comparable to or smaller than the galaxies we wish to study, hence removing them without impacting measurements of galaxy flux is extremely difficult. 

\subsection{Surface photometry}

We use \textsc{sigma} (Structural Investigation of Galaxies via Model Analysis; \citealt{kelvin12}), part of the GAMA pipeline, to calculate S\'{e}rsic magnitudes, S\'{e}rsic indices and half-light radii. \textsc{SIGMA} is a wrapper written in the \textsc{R} programming language \citep{r10} around \textsc{SExtractor}, \citep{bertin96} \textsc{PSFex} \citep{bertin11} and \textsc{galfit} 3 \citep{peng10}. Given the RA and DEC of a GAMA catalogued galaxy, \textsc{SIGMA} calculates its pixel position in the mosaic and produces a 1200 pixel square cutout centered on that galaxy. \textsc{sigma} invokes \textsc{SExtractor} to run on this cutout, producing a catalogue of stars and galaxies which are fed into \textsc{PSFEx}. \textsc{PSFEx} is a point spread function (PSF) \textbf{extraction tool and is used here} to generate an empirical PSF for the cutout. 

\begin{figure}
\begin{center}
\includegraphics[width=3in, angle=0]{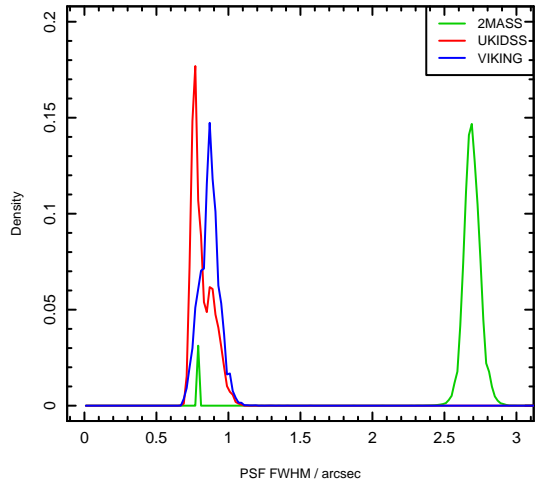}
\caption{[Figure updated]. Histogram of measured PSF FWHMs in 0.02 arcsec bins for 2MASS, UKIDSS-LAS and VISTA VIKING as indicated.}
\label{fig:psf}
\end{center}
\end{figure}

Figure \ref{fig:psf} shows the distribution of PSF FWHMs for the three surveys extracted by \textsc{PSFEx}. The 2MASS, UKIDSS-LAS and VISTA VIKING surveys achieve $3\sigma$ clipped median FWHMs of $2.69 \pm 0.05$, \textbf{$0.80 \pm 0.07$} and $0.87 \pm 0.07$ arcsec respectively. The spike at 0.8 for the 2MASS data is the result of fitting failures.

\begin{sloppypar}
The PSF is used in a second \textsc{SExtractor} run to calculate \textbf{the sky rms and} an object's Kron-like (AUTO) magnitude and radius, half-light radius, ellipticity and CLASS\_STAR probability, which are used as inputs to \textsc{galfit}. \textsc{galfit} is a 2D analysis algorithm capable of fitting multiple common astronomical profiles (e.g. Gaussian, S\'{e}rsic, exponential) to one or more objects with one or more components each within a single image frame. Within \textsc{sigma}, \textsc{galfit} is \textbf{used to simultaneously fit a single S\'{e}rsic profile to the target galaxy and scaled PSFs or S\'{e}rsic profiles to secondary sources as appropriate. \textsc{galfit} does not estimate Poisson noise from the sources in question, however this should not be a issue for ground-based NIR observations.}
\end{sloppypar}

We ran \textsc{sigma} on our SDSS $r < 19.8$~mag defined sample of \textbf{37,591} G09 objects that have a SURVEY\_CLASS $\geq 2$ as defined by the GAMA database (\citealp{driver11}, their Table 4). These objects are derived from SDSS DR8 and have passed GAMA's standard star/galaxy separation criteria (for details, see \citealt{baldry10}). We constrain ellipticity\footnote{$e = 1 - b/a$, where $a=$~semi-major axis, and $b=$~semi-minor axis} to $0 < e < 0.95$ and S\'{e}rsic indices to $0.3 < n < 15$. In calculating S\'{e}rsic magnitudes, we integrate the S\'{e}rsic function (equation \ref{eq:sersic}) to $10 r_e$; this avoids regions where little is known about galaxy light profiles. 

\begin{figure}
\begin{center}
\includegraphics[width=3in, angle=0]{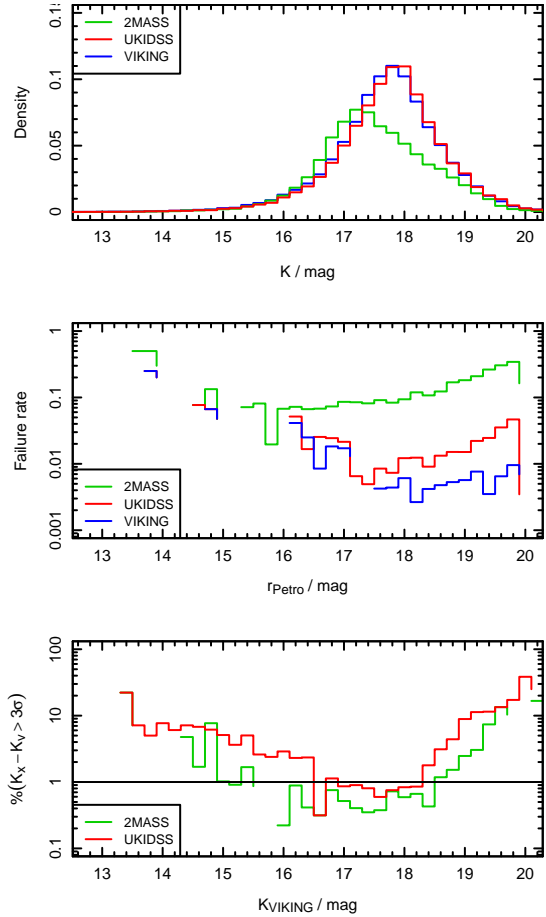}
\caption{[Figure updated]. The distribution of galaxies as a function of the native S\'{e}rsic magnitude (top) and failed fits in the G09 sample as a function of SDSS $r_\mathrm{Petrosian}$ (middle) in intervals of $\Delta m = 0.2$~mag; green: 2MASS, red: UKIDSS-LAS, blue: VISTA VIKING. \textbf{The bottom panel shows the relative fraction of objects where the flux uncertainty exceeds 3$\sigma$. The solid line represents the expected number ($\sim 1\%$) of said outliers.}}
\label{fig:fail}
\end{center}
\end{figure}

\begin{sloppypar}
\textsc{sigma} failed to converge for \textbf{8849 (23.5\%), 1037 (2.8\%) and 248 (0.7\%)} objects in the 2MASS, UKIDSS-LAS and VISTA VIKING data sets respectively. Figure \ref{fig:fail} shows the distribution of the entire sample as a function of the native S\'{e}rsic magnitude excluding convergence failures (upper panel) and objects for which fitting failed with respect to the input SDSS magnitude (middle panel). The increased failure rate at brighter magnitudes may be indicative of a deblending issue exacerbated by small number statistics; for $K_s < 14$~mag there are no more than 40 galaxies per bin. The 2MASS data shows an elevated failure rate \textbf{for $K > 16.5$~mag, increasing from 10\% to 30\%. The VISTA VIKING data performs notably better to the UKIDSS-LAS data over most of the sample with a failure rate of 0 -- 4\% compared to 0.05 -- 5\% and demonstrates an improvement for objects with $r > 17.2$~mag. The failure rate in UKIDSS-LAS increases beyond 1\% at $r = 18.9$~mag while VISTA VIKING shows no such increase within the sample}. 

\end{sloppypar}

\textbf{The bottom panel of Figure \ref{fig:fail} shows the number of 3$\sigma$ outliers, where $\sigma$ is the standard deviation of $K_x - K_v$ (where $x$ = 2MASS, UKIDSS-LAS) in each bin. The fitting failures discussed above have been excluded from this panel. A likely explanation for the increase in outliers at the bright end is the detection of previously undetectable disks in VISTA VIKING. The two surveys perform similarly, with 2MASS performing slightly better --- the number of 3$\sigma$ outliers rises above the expected 1\% threshold at 18.5~mag for 2MASS compared to 18.3~mag for UKIDSS-LAS. However, the random uncertainty in 2MASS magnitudes for $K > 15$~mag is at least 2.5 times larger than UKIDSS-LAS. Furthermore, the number of objects where \textsc{sigma} fails to converge is over 10 times greater for 2MASS than UKIDSS-LAS for $K > 17.2$~mag. }

\section{ROBUSTNESS OF FLUX AND STRUCTURE MEASUREMENTS}
\label{sec:results}

\subsection{Star-galaxy separation limits}

\begin{figure}
\begin{center}
\includegraphics[width=3in, angle=0]{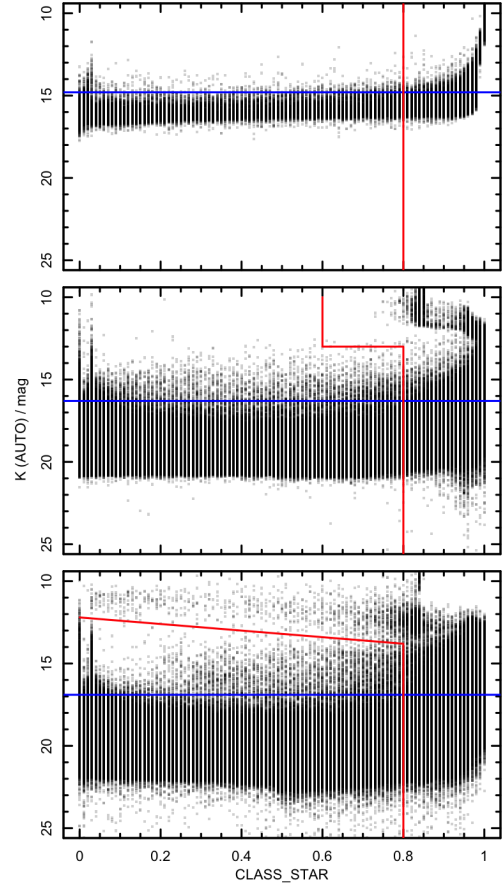}
\caption{Probability of being a star (CLASS\_STAR) vs \textsc{SExtractor}'s Kron-like (AUTO) $K_s$ band magnitude; top: 2MASS, middle: UKIDSS-LAS, bottom: VISTA VIKING. The red lines illustrate star-galaxy separation criteria; items to the left and below the red line are considered to be galaxies.}
\label{fig:classification}
\end{center}
\end{figure}

Figure \ref{fig:classification} shows \textsc{SExtractor}'s CLASS\_STAR probability versus the $K_s$ band AUTO magnitude. We find that we can unambiguously distinguish between galaxies and stars to 14.8~mag, 16.3~mag and 16.9~mag for 2MASS, UKIDSS-LAS and VISTA VIKING respectively, as indicated by the blue horizontal line. However, star-galaxy separation performs poorly in the VISTA VIKING data set for very bright objects due to saturated stars with $K_s \leq 13$~mag, illustrating that increased sensitivity typically comes with the drawback of a fainter saturation magnitude if the exposure time remains constant. The red lines in the Figure show our star-galaxy separation criteria. We generally consider objects with CLASS\_STAR $< 0.8$ to be galaxies however, to exclude saturated stars, we reject UKIDSS objects that do not satisfy CLASS\_STAR $< 0.6$ and $K_s > 13$~mag. We reject VISTA VIKING objects with $K_s - 2~\mathrm{CLASS\_STAR}  < 12.2$~mag for the same reason.

\subsection{Source detection}

For source extraction, we run \textsc{SExtractor} on the mosaics generated by \textsc{swarp}. \textsc{SExtractor} is a program that builds catalogues of sources, their positions and Petrosian and Kron-like fluxes from an input image. We use a detection threshold of $2\sigma$ above the background, a minimum object size (\textsc{SExtr\-actor} parameter DETECT\_MIN\-AREA) of 175, 7 and 5 pixels to maintain a detection area threshold of 5 native pixels and a seeing FWHM of 2.69, 0.80 and 0.87 arcsec for 2MASS, UKIDSS-LAS and VISTA VIKING respectively. We select objects with \textsc{SExtractor} FLAGS $< 4$ to ensure reliable photometry. This resulted in a total of \textbf{54,412, 558,748 and 1,438,461} sources for 2MASS, UKIDSS-LAS and VISTA VIKING respectively.

\label{sec:detect}

\begin{figure}
\begin{center}
\includegraphics[width=3in, angle=0]{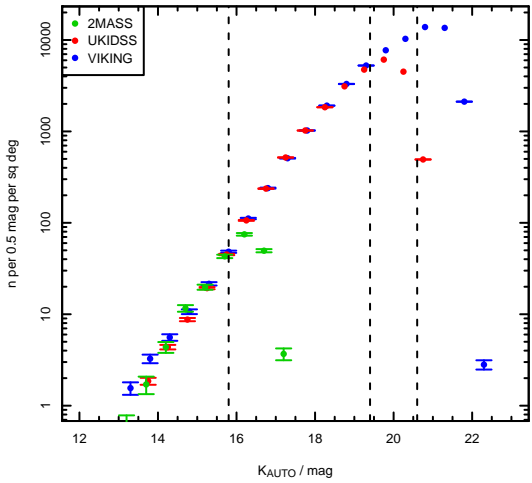}
\caption{Galaxy counts as a function of magnitude, binned in $\Delta m = 0.5$~mag intervals.}
\label{fig:galmag}
\end{center}
\end{figure}

Figure \ref{fig:galmag} shows galaxy number counts as a function of the $K_s$ band AUTO magnitude for the three data sets. The distributions peak at 16.3~mag, 19.9~mag and 21.1~mag for the 2MASS, UKIDSS-LAS and VISTA VIKING data sets respectively. However, in each case, the distribution deviates slightly from the observed exponential increase due to incompleteness, presumably caused by extended or distant objects falling below the isophotal detection threshold. To account for this effect, we set our inferred detection limits (indicated by the vertical dashed lines) 0.5~mag brighter; hence our nominal $2\sigma$ detection limits for 2MASS, UKI\-DSS-LAS and VISTA VIKING are 15.8~mag, 19.4~mag and 20.6~mag respectively. We note the VISTA VIKING data is sensitive enough to detect the change in logarithmic slope at $K_s \sim 19$~mag (see \citealt{christobal09}). G09 is 29\% underdense out to $z < 0.1$ \citep{driver11}, therefore there are fewer bright objects than expected.

\subsection{Surface brightness limits}

\begin{figure}
\begin{center}
\includegraphics[width=3in]{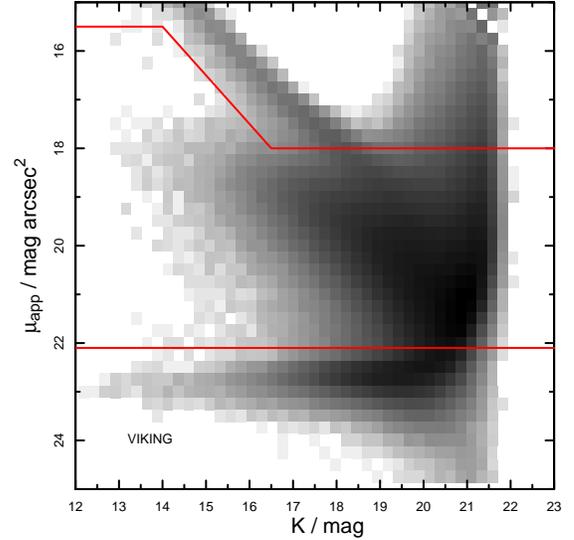}
\caption{Raw apparent surface brightness as a function of magnitude of all VISTA VIKING sources, with density scaling logarithmically from sparsest (white) to densest (black). Bin sizes are 0.25 $\times$ 0.25.}
\label{fig:bbd}
\end{center}
\end{figure}

\begin{sloppypar}
Figure \ref{fig:bbd} shows the raw apparent surface brightness/apparent magnitude relation of sources detected in the VISTA VIKING dataset. We calculate the effective (or half-light) surface brightness of an object using equation A1 of \citet{driver05}: 

\begin{equation}
\mu_\mathrm{app} = K_s + 2.5 \log {2 \pi (r_e^2 - 0.32 \Gamma^2)}
\end{equation}
\\
where $K_s$ is the $K_s$ band AUTO magnitude, $r_e$ is the \textsc{SExtractor} derived half-light radius (FLUX\_RADIUS) in arcsec and $\Gamma$ is the seeing FWHM peak from Figure \ref{fig:psf}.  The diagonal feature in the top center represent stars that were not excluded by our CLASS\_STAR cuts. This is because we have assumed a constant mean seeing across the mosaic. Some contributing frames will have an FWHM less than this, leading to stars potentially being misclassified. Objects to the right and above of the stellar locus are cosmic rays and noise, while the horizontal streak at $22 < \mu_{app} < 23.5$~mag~arcsec$^{-2}$ represents noise detections at the surface brightness limit. To exclude these in number counts, we require sources to have $\mu_{app} < 22.1$~mag~arcsec$^{-2}$ and $\mu_{app} > 15.5, K_s - 1.5, 18$~mag~arcsec$^{-2}$ for $K_s < 14$~mag, $14 < K_s < 16.5$~mag and $K_s > 16.5$~mag respectively. These cuts are indicated by the red lines.

\begin{figure}
\begin{center}
\includegraphics[width=3in]{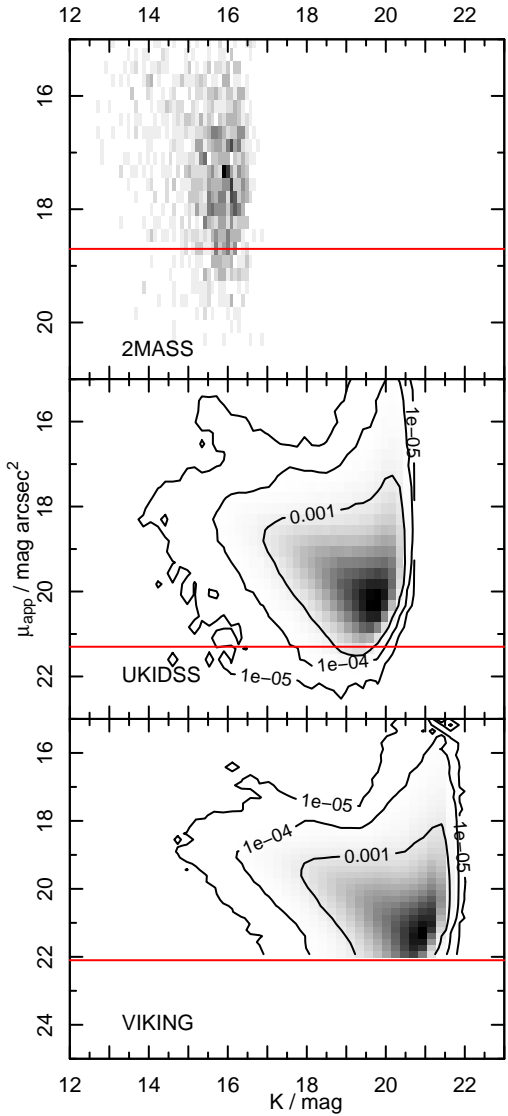}
\caption{Apparent surface brightness as a function of magnitude for all three data sets, with density scaling linearly from sparsest (white) to densest (black). The red horizontal lines represent the surface brightness detection threshold. The contours contain 99.999, 99.99 and 99.9 per cent of the data respectively. Bin sizes are 0.25 $\times$ 0.25.}
\label{fig:bbd3}
\end{center}
\end{figure}

Figure \ref{fig:bbd3} shows the apparent surface brightness/apparent magnitude relation of galaxies for the three data sets. Both UKIDSS-LAS and 2MASS data sets have been matched against the cleaned VISTA VIKING data set and for $K < 18$~mag, we require a source to be detected as a galaxy in both UKIDSS-LAS and VISTA VIKING. We note that \textsc{SExtractor} half-light radii are overestimated using the 2MASS dataset by a median factor of 1.94, presumably due to low resolution and smearing. This is corrected for in the Figure.  The apparent surface brightness limits are 19.9~mag arcsec$^{-2}$, 21.9~mag arcsec$^{-2}$ and 22.1 mag arcsec$^{-2}$ for 2MASS, UKIDSS-LAS and VISTA VIKING respectively. These will, of course, translate to brighter absolute surface brightness limits depending on each object's individual redshift. 
\end{sloppypar}

\subsection{Photometric accuracy limits}


\begin{figure}
\begin{center}
\includegraphics[width=3in]{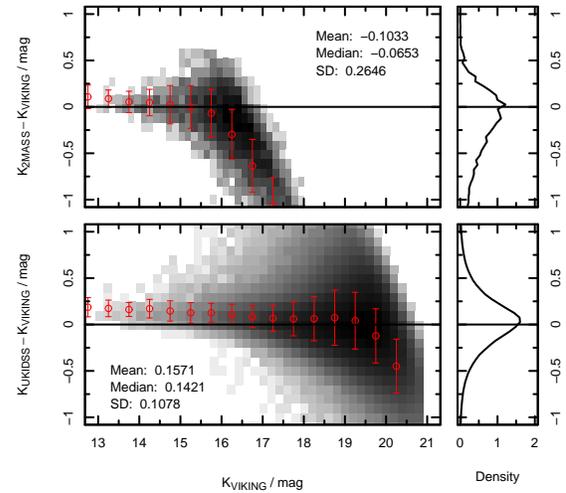}
\caption{2D histogram of the differences between the AUTO magnitudes derived from the 2MASS (top) and UKIDSS-LAS (bottom) datasets and VISTA VIKING as a function of the VISTA VIKING derived AUTO magnitude with $3 \sigma$ clipped statistics for $13 < K_s < 16$ inset. Density varies logarithmically from white (sparsest) to black (densest). Bin sizes are 0.2~mag $\times$ 0.1~mag. The histograms on the right show counts as a function of $\Delta m$ with bin size 0.05. The error bars show the median and standard deviation in each 0.5~mag interval.}
\label{fig:trumpet}
\end{center}
\end{figure}

Figure \ref{fig:trumpet} compares the consistency of \textsc{SExtractor} photometry derived from the 2MASS and UKIDSS-LAS datasets relative to the VISTA VIKING data set. To calibrate for the difference between the 2MASS and UKIDSS versions of the $K_s$ filters, we use the colour equations and the $H_\mathrm{AB} - K_\mathrm{AB}$ offset of $-0.5$~mag in \citet{hewitt06} and a typical $H_\mathrm{Vega} - K_\mathrm{Vega} \simeq 0.25$~mag \citep{bell03}.  


The 2MASS dataset struggles for objects fainter than $K_s = 16$~mag, where objects are consistently brighter compared to VISTA VIKING presumably due to low resolution and noise. Despite this, the 2MASS dataset shows agreement to $\pm$0.1~mag and $\pm$0.2~mag (1 $\sigma$, $3 \sigma$ clipped) to $K_s = 13.1$~mag and 15.3~mag respectively. The corresponding limits for UKIDSS-LAS are 16.3~mag and 18.3~mag. There appears to be a non-trivial ($\sim0.15$~mag) offset between UKIDSS-LAS and VISTA VIKING magnitudes, which decreases as a function of magnitude. A similar, but not statistically significant offset is observed when comparing S\'{e}rsic magnitudes between UKIDSS-LAS and VISTA VIKING for the SDSS sample. The cause of this offset and its variance as a function of magnitude is currently under investigation with one suggested cause being linearity issues with the calibration of at least one of the detectors (Driver et al., in prep).

\begin{table}
\begin{center}
\caption{Maximum precision and $1 \sigma$ limiting magnitudes for the given $\Delta m$.}
\label{table:trumpet}
\begin{tabular}{| c || c | c | c |}
\hline
Survey & Min               & \multicolumn{2}{|c|}{Mag limit for:} \\ \cline{3-4}
              & $\Delta m$  &  $\leq 0.1$ mag & $\leq 0.2$ mag \\
\hline
2MASS    & 0.05 & 13.1 & 15.3 \\
UKIDSS   & 0.07 & 16.3 & 18.3 \\
\hline
\end{tabular}
\end{center}
\end{table}

\subsection{Structural accuracy limits}
\subsubsection{Half-light radius}

\begin{figure}
\begin{center}
\includegraphics[width=3in, angle=0]{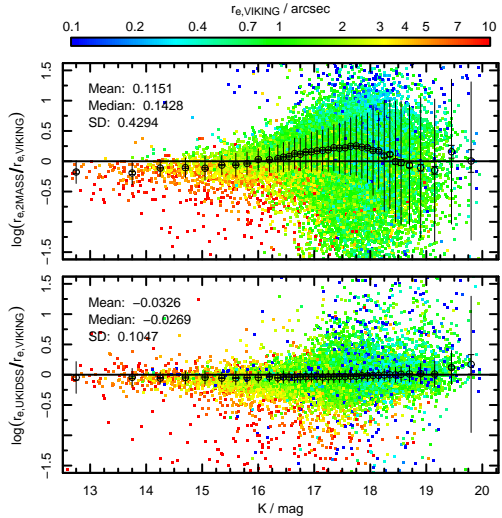}
\caption{Ratio of half-light radii as a function of the VISTA VIKING $K_s$ magnitude. Points are coloured according to their VISTA VIKING half-light radius. The black points, lines and error bars show $3 \sigma$ clipped medians, standard deviations and standard errors plotted at bin midpoints.}
\label{fig:recompare}
\end{center}
\end{figure}

Figure \ref{fig:recompare} shows the ratio of half-light radii derived from \textsc{sigma} photometry compared to the VIKING half-light radius. Although galaxies appear to be bigger as depth increases (see Figure \ref{fig:pictures}), this is a visual illusion caused by the fainter background. In fact, the effective radius from UKIDSS-LAS to VISTA VIKING stays constant with the median deviation being statistically insignificant over entire sample: $-0.027 \pm 0.105$ ($3 \sigma$ clipped). The half-light radius from 2MASS to VISTA VIKING also shows a small deviation in the S\'{e}rsic parameters, with a median deviation of $0.14 \pm 0.43$. This suggests the S\'{e}rsic fits are reasonable, and the increased apparent size is merely a result of decreased background and noise. Considering these caveats, the half-light radius is surprisingly robust: both 2MASS and UKIDSS half-light radii are within 0.2~dex of VIKING up to 16.1~mag and 19.0~mag respectively. Table \ref{table:re} shows the limiting magnitude for errors of 0.1~dex, 0.2~dex and 0.3~dex.

\begin{table}
\begin{center}
\caption{Limiting magnitudes for the given precision (in dex) in half-light radii.}
\label{table:re}
\begin{tabular}{| c || c | c | c | c |}
\hline
Survey & Min               & \multicolumn{3}{|c|}{Mag limit for:} \\ \cline{3-5}
              & $\Delta r_e$  &  $\leq 0.1$ & $\leq 0.2$ & $\leq 0.3$ \\ 
\hline
2MASS  & 0.16 & ---     & 16.1 & 16.9 \\
UKIDSS-LAS & 0.09 & 18.3 & 19.0 & 19.6 \\
\hline
\end{tabular}
\end{center}
\end{table} 

\subsubsection{Ellipticity}

\begin{figure}
\begin{center}
\includegraphics[width=3in, angle=0]{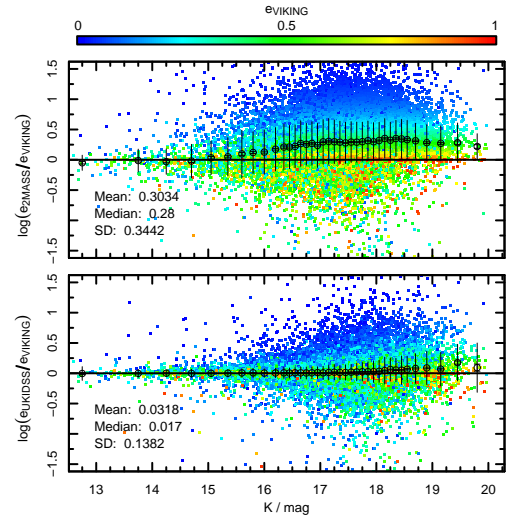}
\caption{Ratio of ellipticities as a function of the VISTA VIKING $K_s$ magnitude. Points are coloured according to their VISTA VIKING ellipticity. The black points, lines and error bars show $3 \sigma$ clipped medians, standard deviations and standard errors plotted at bin midpoints.}
\label{fig:ellipcompare}
\end{center}
\end{figure}

Figure \ref{fig:ellipcompare} shows the ratio of \textsc{sigma} derived ellipticities relative to the VISTA VIKING data set. It should not be a surprise that objects with low ellipticity tend to have large fluctuations between data sets; this is due to the greater relative error. Beyond $K_s > 15$~mag, 2MASS galaxies appear to be consistently more elliptical than the VISTA VIKING data, presumably because of a combination of noise and low resolution. For this reason 2MASS data performs rather poorly --- errors increase to $>0.2$~dex at $K_s = 14.0$~mag. UKIDSS-LAS is within 0.2~dex of VISTA VIKING ellipticities down 18.4~mag. Table \ref{table:ellip} shows the limiting magnitude corresponding to the given precisions.

\begin{table}
\begin{center}
\caption{Limiting magnitudes for the given precision (in dex) in ellipticity.}
\label{table:ellip}
\begin{tabular}{| c || c | c | c | c |}
\hline
Survey & Min               & \multicolumn{3}{|c|}{Mag limit for:} \\ \cline{3-5}
              & $\Delta e$  &  $\leq 0.1$ & $\leq 0.2$ & $\leq 0.3$ \\ \hline
2MASS  & 0.12 & ---     & 14.0 & 15.1 \\
UKIDSS-LAS & 0.06 & 17.0 & 18.4 & 19.3 \\
\hline
\end{tabular}
\end{center}
\end{table} 

\subsubsection{S\'{e}rsic index}

\begin{figure}
\begin{center}
\includegraphics[width=3in, angle=0]{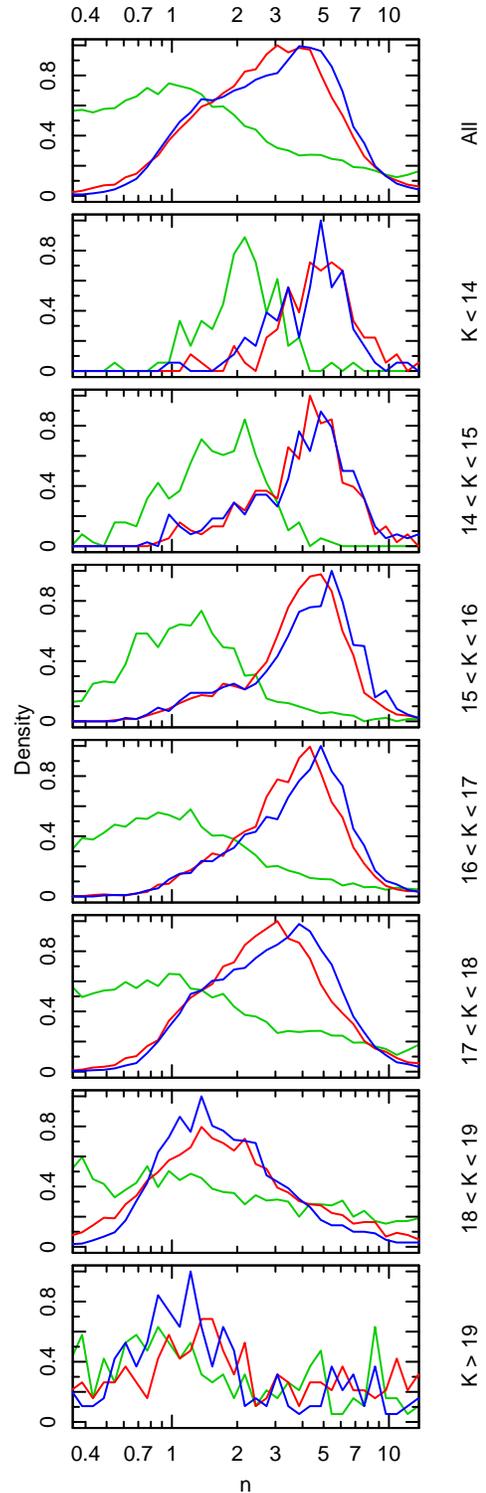}
\caption{Histograms of S\'{e}rsic indices in the given S\'{e}rsic magnitude ranges for the three data sets convolved with a rectangular kernel of width $\log n = 0.05$; green: 2MASS, red: UKIDSS-LAS, blue: VISTA VIKING.}
\label{fig:sersichisto}
\end{center}
\end{figure}

\begin{figure}
\begin{center}
\includegraphics[width=3in, angle=0]{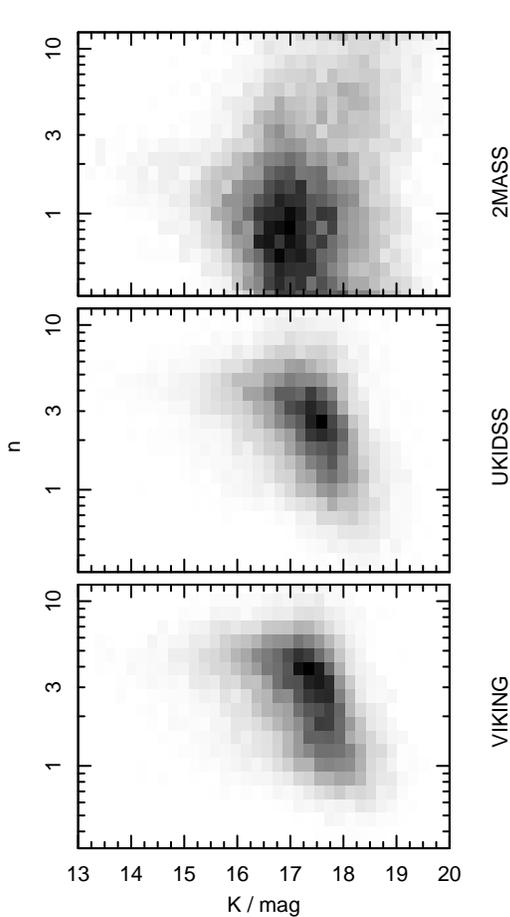}
\caption{2D histogram of the distribution of recovered S\'{e}rsic indices as a function of $K_s$ linearly ranging from white (sparsest) to black (densest), with bin sizes of 0.2~mag $\times$ 0.08 in $\log n$. }
\label{fig:logn}
\end{center}
\end{figure}

\begin{figure}
\begin{center}
\includegraphics[width=3in, angle=0]{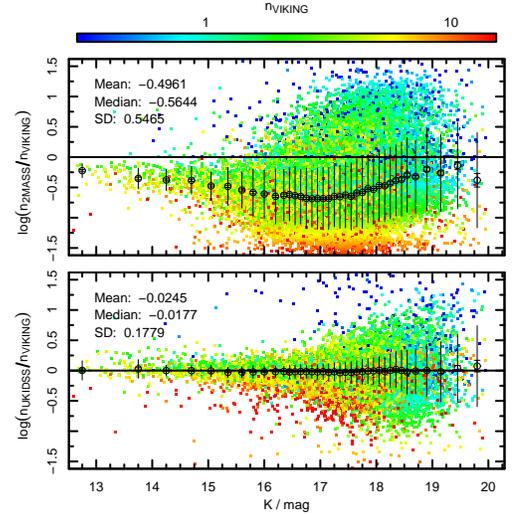}
\caption{Ratio of S\'{e}rsic indices as a function of the VISTA VIKING derived S\'{e}rsic magnitude. Points are coloured according to their VISTA VIKING derived S\'{e}rsic index. The black points, lines and error bars show $3 \sigma$ clipped medians, standard deviations and standard errors plotted at bin midpoints.}
\label{fig:sersiccompare}
\end{center}
\end{figure}

Figure \ref{fig:sersichisto} presents a histogram of the recovered S\'{e}rsic indices for each data set. The number of objects with the Gaussian $n = 0.5$ profile declines from 2MASS to UKIDSS-LAS then VISTA VIKING, indicating a greater number of objects are resolved in the latter data sets, particularly at the fainter magnitudes. \citet{kelvin12} find that the distribution of S\'{e}rsic indices in optical wavelengths is bimodal, resembling the sum of Gaussian distributions centered on $n \sim 1$ and $n \sim 3.5$ which correspond to late and early type galaxies respectively. The two peaks blend with increasing wavelength --- NIR data is a more effective probe of older stellar populations that dominate early type galaxies and bulges while optical data tracks younger stellar populations more prevalent in late type galaxies. Interestingly, it appears that using the deeper VISTA VIKING dataset shifts the peaks outwards towards the exponential $n = 1$ and de Vaucouleurs $n = 4$ profiles. In particular, there are relatively more high S\'{e}rsic index objects at $15 < K_s < 18$. This is a consequence of increased S/N and hence more robust sky modelling in the VIKING data set (see Figures \ref{fig:pictures} and \ref{fig:logn}). Figure \ref{fig:logn} demonstrates that using VISTA VIKING over UKIDSS-LAS does not result in the detection of objects at fainter magnitudes for a given S\'{e}rsic index, this may reflect a shortcoming of our sample. The 2MASS dataset does not display the double-peaked distribution at any magnitude within the sample presumably due to the lack of resolution; furthermore the bright sky makes disk ($n = 1$) detection difficult.

Figure \ref{fig:sersiccompare} shows the ratio of S\'{e}rsic indices as a function of the S\'{e}rsic magnitude. It is clear from the Figure that the 2MASS dataset is \textbf{useless} even for the brightest ($K_s < 14$~mag) galaxies, consistently underestimating the S\'{e}rsic index due to lack of resolution. At fainter magnitudes, galaxies in the sample transition toward becoming unresolved point sources, and hence having a Gaussian profile --- in Figure \ref{fig:sersichisto} we see the distribution peaking at lower S\'{e}rsic indices with fainter magnitudes. This transition occurs at $K_s \sim 17$~mag for both the UKIDSS-LAS and VISTA VIKING datasets. There is no systematic offset when comparing UKIDSS-LAS against VISTA \textbf{VIKING. We find the S\'{e}rsic indices are remarkably robust, being consistent to 0.2~dex down to 17.7~mag}. Table \ref{table:sersic} shows limiting magnitudes for the given precisions.

\begin{table}
\begin{center}
\caption{Limiting magnitudes for the given precision (in dex) in S\'{e}rsic index.}
\label{table:sersic}
\begin{tabular}{| c || c | c | c | c |}
\hline
Survey & Min               & \multicolumn{3}{|c|}{Mag limit for:} \\ \cline{3-5}
              & $\Delta n$  &  $\leq 0.1$ & $\leq 0.2$ & $\leq 0.3$ \\ \hline
2MASS  & ---  & ---     & ---  & ---  \\
UKIDSS-LAS & 0.08 & 16.4 & 17.7 & 18.4 \\
\hline
\end{tabular}
\end{center}
\end{table} 

\section{SUMMARY}
\begin{table*}[h!]
\begin{center}
\caption{Summary of results: limiting magnitudes and cumulative densities per deg$^2$}
\label{table:numbers}
\begin{tabular}{| c || c | c | c | c | c | c |}
\hline Limits & \multicolumn{2}{|c|}{2MASS} & \multicolumn{2}{|c|}{UKIDSS} & \multicolumn{2}{|c|}{VIKING$^a$} \\
\hline & Limit & Density$^b$ & Limit & Density$^b$ & Limit & Density$^b$ \\
\hline
PSF FWHM (arcsec) & 2.7 & --- & 0.80 & ---  & 0.87 & ---      \\
Star-galaxy separation limit$^d$ & 14.8 & 18 & 16.3 & 130 & 16.9 & 280 \\
Detection limit (\textit{r}, \textbf{1}\% failure) & --- & --- & 18.9 & --- & $> 19.8$ & --- \\
Surface brightness$^e$ & 18.6 & ---   & 21.3 &     ---   & 22.1   & ---          \\
Counts$^d$                      & 15.8 &   70 & 19.4 & 6700 & 20.6   & 21000 \\


Photometric limit ($\Delta m \leq 0.2$ mag)$^c$    & 14.7 & 12  & 18.5 & 1500 & (19.6) &     4900 \\
S\'{e}rsic index ($\Delta n \leq 0.2$ dex) $^c$ & --- & ---  & 17.7 & 540 & (18.8) & 2100  \\
Half-light radius ($\Delta r_e \leq 0.2$ dex) $^c$ & 18.3 & 1100 & 19.0 & 2600 & (20.1) & 8400  \\
Ellipticity ($\Delta e \leq 0.2$ dex) $^c$ & 14.0 & 5.6  & 18.4 & 1300 & (19.5) &   4400   \\
\hline
\end{tabular}
\medskip\\
$^a$Bracketed values inferred from UKIDSS-LAS by adding 1.1~mag \\
$^b$Cumulative densities in N per mag per deg$^2$ \\
$^c$S\'{e}rsic magnitude \\
$^d$AUTO magnitude \\
$^e$AUTO magnitude arcsec$^{-2}$\\
\end{center}
\end{table*}

\begin{sloppypar}
Throughout this paper, we have demonstrated the benefits of applying galaxy photometry and profiling to more sensitive data sets. Table \ref{table:numbers} summarizes our findings. We have constructed renormalized mosaics of 2MASS, UKIDSS-LAS and VISTA VIKING imagery of the G09 region. Using these mosaics, we establish in Figures \ref{fig:pictures} and \ref{fig:colorpictures} that the VISTA VIKING data is visually superior, with a higher S/N and lower sky background. 
\end{sloppypar}

We apply \textsc{SExtractor}'s star-galaxy separation to the three data sets, finding that using the VISTA VIKING data we are able to separate stars and galaxies to 0.5~mag fainter than UKIDSS-LAS and 1.5~mag fainter than 2MASS (see Figure \ref{fig:classification}). We caution that star-galaxy separation performs poorly at brighter magnitudes ($K_s < 14$~mag) in the VISTA VIKING data and, to a lesser extent, UKIDSS-LAS due to a lower saturation threshold. We establish \textsc{SExtractor} photometric limits of 15.8, 19.4 and 20.6 mag, \textbf{star-galaxy separation limits of 14.8, 16.3 and 16.9~mag} and surface brightness limits of 18.6, 21.3, and 22.1 mag~arcsec$^{-2}$ for 2MASS, UKIDSS-LAS and VISTA VIKING respectively. We find that there is a $\sim0.15$~mag offset between VISTA VIKING and existing UKIDSS-LAS photometry which is currently under investigation.

We used \textsc{sigma} to fit S\'{e}rsic profiles to a sample of \textbf{37,591} SDSS galaxies with $r_\mathrm{Petrosian} \leq 19.8$~mag. Using the VISTA VIKING data set yields a decrease in the number of convergence failures compared to UKIDSS-LAS for $r > 17.3$~mag. The VISTA VIKING failure rate does not increase appreciably at the fainter end of the sample. The effective radius is the most stable of the S\'{e}rsic parameters, with the 2MASS and UKIDSS-LAS data being within 0.2~dex of the VISTA VIKING data down to 16.1~mag and 19.0~mag respectively. The 2MASS dataset performs poorly over most of the sample when recovering ellipticities, with the deviation from VISTA VIKING rising above 0.2~dex at $K_s = 14.0$~mag. UKIDSS-LAS derived ellipticities are more stable, where the deviation increases beyond 0.2~dex at $K_s = 18.4$~mag. The 2MASS dataset yields a systematic offset when compared to VISTA VIKING in the S\'{e}rsic index for the entire sample. 

 
\begin{acknowledgements}
We thank the anonymous referee whose comments helped improve this paper.

SKA is supported by a summer scholarship provided by the School of Physics at UWA. LSK is supported by the Austrian Science Foundation FWF under grant P23946.

\begin{sloppypar}
GAMA is a joint European-Australasian project based around a spectroscopic campaign using the Anglo-Australian Telescope. The GAMA input catalogue is based on data taken from the Sloan Digital Sky Survey and the UKIRT Infrared Deep Sky Survey. Complementary imaging of the GAMA regions is being obtained by a number of independent survey programs including \textit{GALEX} MIS, VST KIDS, VISTA VIKING, \textit{WISE}, \textit{Herschel}-ATLAS, GMRT and ASKAP providing UV to radio coverage. GAMA is funded by the STFC (UK), the ARC (Australia), the AAO, and the participating institutions. The GAMA website is \url{http://www.gama-survey.org}. 
\end{sloppypar}

We gratefully acknowledge use of data from the ESO Public Survey programme ID 179.B-2004 taken with the VISTA telescope, data products from CASU and VSA archive operated by WFAU.

This work is based in part on data obtained as part of the UKIRT Infrared Deep Sky Survey. The UKIDSS project is defined in \citet{lawrence07}. UKIDSS uses the UKIRT Wide Field Camera (WFCAM; \citealt{casali07}). The photometric system is described in \citet{hewitt06}, and the calibration is described in \citet{hodgekin09}. The pipeline processing and science archive are described in Irwin et al (2009, in prep) and \citet{hambly08}.

This publication makes use of data products from the Two Micron All Sky Survey, which is a joint project of the University of Massachusetts and the Infrared Processing and Analysis Center/California Institute of Technology, funded by the National Aeronautics and Space Administration and the National Science Foundation.
\end{acknowledgements}

\begin{appendix}

\section{APPENDIX: DESCRIPTION OF CATALOGUES}

\textbf{We release two catalogues with this paper. The first, \texttt{2uvsextract-trimmed.fits}, contains the output from \texttt{SExtractor} and is described in Table \ref{table:cat1}. Bad or missing values are blank. \texttt{2uvsigma-trimmed.fits} contains the output from \textsc{sigma} and is described in Table \ref{table:cat2}. Coverage of the entire G09 region may be incomplete for all three surveys. Missing values are denoted by \texttt{-9999}. The suffixes \texttt{\_2}, \texttt{\_U} and \texttt{\_V} denote 2MASS, UKIDSS-LAS and VISTA VIKING. The prefix \texttt{M01\_} denotes quantities measured by \textsc{galfit}. These catalogues are available from \url{http://star-www.st-and.ac.uk/~spd3/2uvsextractor-trimmed.fits} and \url{http://star-www.st-and.ac.uk/~spd3/2uvsigma-trimmed.fits}.}

\begin{table}
\begin{center}
\caption{Description of \texttt{2uvsextract.fits}.}
\label{table:cat1}
\begin{tabular}{ m{2.4cm} m{5.1cm} }
\texttt{NAME}				& Unique identifier \\
\texttt{RA}					& Right ascension (J2000) \\
\texttt{DEC}				& Declination (J2000) \\
\texttt{MAG\_AUTO}			& Kron-like (AUTO) magnitude \\
\texttt{ERRMAG\_AUTO} 		& Error in AUTO magnitude \\
\texttt{FLUX\_RADIUS}		& Half-light radius (pixel) \\
\texttt{KRON\_RADIUS}		& Kron radius used to calculate \texttt{MAG\_AUTO}  (pixel) \\
\texttt{ISOAREAF\_IMAGE}	& Filtered isophotal area above detection threshold (pixel$^2$) \\
\texttt{MU\_THRESHOLD}		& Surface brightness detection threshold (mag~arcsec$^{-2}$) \\
\texttt{MU\_MAX}			& Maximum surface brightness (mag~arcsec$^{-2}$) \\
\texttt{FWHM\_IMAGE}		& PSF FWHM (pixel) \\
\texttt{CLASS\_STAR}		& Probability an object is a star \\
\texttt{FLAGS} 				& \textsc{SExtractor} flags
\end{tabular}
\end{center}
\end{table}

\begin{table*}
\begin{center}
\caption{Description of \texttt{2uvsigma.fits}.}
\label{table:cat2}
\begin{tabular}{ m{4.3cm} m{11.2cm} }
\texttt{CATAID}				& GAMA object ID \\
\texttt{RA}					& Right ascension (J2000) \\
\texttt{DEC}				& Declination (J2000) \\
\texttt{R\_PETRO}			& \textit{r} band Petrosian magnitude \\
\texttt{SURVEY\_CLASS}		& GAMA star/galaxy classification, see \citet{driver11} \\
\texttt{SKY}				& \textsc{SExtractor} sky background level (ADU) \\
\texttt{SKY\_ERR}			& \textsc{SExtractor} error in sky background level (ADU) \\
\texttt{SKY\_RMS} 			& \textsc{SExtractor} RMS of sky background (ADU) \\
\texttt{CENFLUX}			& Central flux of object (ADU) \\
\texttt{PSFNUM}			& Number of sources used to estimate PSF \\
\texttt{PSFCHI2}			& \textsc{PSFEx} PSF $\chi^2$ \\
\texttt{PSFFWHM} 			& \textsc{PSFEx} PSF FWHM (arcsec) \\
\texttt{FLUX\_RADIUS}		& \textsc{SExtractor} half-light radius (arcsec) \\
\texttt{KRON\_RADIUS}		& Kron radius used to calculate \texttt{MAG\_AUTO} (pixel) \\
\texttt{A\_IMAGE}			& \textsc{SExtractor} semi-major axis (arcsec) \\
\texttt{B\_IMAGE} 			& \textsc{SExtractor} semi-minor axis (arcsec) \\
\texttt{MAG\_AUTO}			& Kron-like (AUTO) magnitude \\
\texttt{THETA\_IMAGE} 		& \textsc{SExtractor} position angle \\
\texttt{ELLIPTICITY}			& \textsc{SExtractor} ellipticity \\
\texttt{CLASS\_STAR} 		& \textsc{SExtractor} star/galaxy classifier output \\ 
\texttt{SEXRE}				& \textsc{SExtractor} half-light radius on semi-major axis (arcsec)\\
\texttt{SEXRA}				& \textsc{SExtractor} right ascension (J2000) \\
\texttt{SEXDEC} 			& \textsc{SExtractor} declination (J2000) \\
\texttt{M01\_GALPLAN}		& Number of times \textsc{galfit} was run \\
\texttt{M01\_GALMAG}		& S\'{e}rsic magnitude integrated to infinity \\
\texttt{M01\_GALRE}			& \textsc{galfit} half-light radius along semi-major axis (arcsec) \\
\texttt{M01\_GALINDEX}		& S\'{e}rsic index \\
\texttt{M01\_GALELLIP}		& \textsc{galfit} ellipticity \\
\texttt{M01\_PA}			& \textsc{galfit} position angle \\
\texttt{M01\_GALMAGERR} 	& Error in S\'{e}rsic magnitude \\
\texttt{M01\_GALREERR}		& Error in half-light radius \\
\texttt{M01\_GALINDEXERR}	& Error in S\'{e}rsic index \\
\texttt{M01\_GALELLIPERR} 	& Error in ellipticity \\
\texttt{M01\_PAERR}			& Error in position angle \\
\texttt{M01\_GALRA}			& \textsc{galfit} right ascension (J2000) \\
\texttt{M01\_GALDEC}		& \textsc{galfit} declination (J2000) \\
\texttt{M01\_GALMAG10RE}	& S\'{e}rsic magnitude truncated at $10 r_e$ \\
\texttt{M01\_GALMU0}		& Central surface brightness (mag~arcsec$^{-2}$) \\
\texttt{M01\_GALMUE} 		& Effective surface brightness on semi-major axis (mag~arcsec$^{-2}$)\\
\texttt{M01\_GALMUEAVG}	& 2D Effective surface brightness (mag~arcsec$^{-2}$) \\
\texttt{M01\_GALR90} 		& Radius containing 90\% of light on semi-major axis (arcsec) \\
\texttt{M01\_GALCHI2FULL}	& \textsc{galfit} $\chi^2$ \\
\texttt{M01\_GALNDOF} 		& \textsc{galfit} model degrees of freedom \\
\texttt{M01\_GALCHI2}		& \textsc{galfit} reduced $\chi^2$ \\
\texttt{M01\_PRICHI2FULL}	& $\chi^2$ of primary galaxy \\
\texttt{M01\_PRINDOF}		& Degrees of freedom for primary galaxy \\
\texttt{M01\_PRICHI2} 		& Reduced $\chi^2$ for primary galaxy \\
\texttt{M01\_PRINFP}		& Number of free parameters for primary galaxy \\ 
\end{tabular}
\end{center}
\end{table*}

\end{appendix}

\end{document}